\documentclass{aa} 

\usepackage{graphicx}
\usepackage{txfonts}
\usepackage{threeparttable}
\usepackage{amsmath}
\usepackage{mathtools}

\usepackage{subfigure} 
\usepackage[colorlinks=true, allcolors=blue]{hyperref}
\newif\ifhidealgebra
\hidealgebratrue

\begin{document} 

    \title{Peering through the veil: Investigating protoplanetary disk outer edges using backside visibility}
    \titlerunning{Visibility of protoplanetary disk backsides.}

   \author{J. George \inst{1}
   \and C. Dominik \inst{1}
   \and C. Ginski \inst{2}}
   
   \institute{Anton Pannekoek Institute for Astronomy, University of Amsterdam, Science Park 904, 1098 XH Amsterdam, The Netherlands
    \email{joelgmodiyil@gmail.com} and
           \email{dominik@uva.nl}
    \and
    School of Natural Sciences, Center for Astronomy, University of Galway, Galway, H91 CF50, Ireland
    }
   \date{Received March 11, 2025; accepted June 2, 2025}

\abstract
   {Protoplanetary disks observed in scattered light reveal essential insights into the disk’s 3D architecture and dust properties. These disks, which play a crucial role in planet formation, have complex structures where the visibility of the far side of the disk can vary significantly based on several parameters.}
   {This study aims to explore the factors impacting backside visibility in protoplanetary disks, particularly under variations in inclination, dust distribution, grain characteristics, and outer disk morphology.}
   {Using radiative transfer simulations, we investigate how these variables influence the appearance of the backside in scattered light images.}
   {Tapered disk models with exponential tapers often obscure the backside, which supports the rarity of observed backside features. In cases where backside features are visible at lower inclinations, they likely indicate cut-off disks, as backside detection is challenging in standard tapered models at these inclinations. Additionally, factors such as dust mass, grain distribution, and disk material stratification play crucial roles in backside observability, affecting its potential detection in real observations.}
   {This study contributes to understanding the detectability of the backside in protoplanetary disks, with implications for refining observational strategies and interpreting backside features in scattered light images. These findings help frame backside visibility as a critical aspect of assessing disk structure and evolution.}


   \maketitle

\section{Introduction}

Protoplanetary disks are the sites of planet formation. Observations of these disks provide critical insights into the mechanisms driving planetary system formation. These observations rely on three primary observational techniques: thermal continuum emission, which traces large dust grains in the midplane; scattered light imaging, which highlights small grains in the surface layers; and spectral line emission, which maps the gas component through molecular transitions. Together, these methods and advanced technologies continue to deepen our understanding of the origins of planetary systems across the cosmos.

Scattered light imaging captures starlight scattered by
micrometre-sized dust grains in the disc's upper layers, offering
critical insights into the 3D disk structure, dust grain properties,
and their interaction with stellar radiation. Observations at optical
and near-infrared wavelengths are sensitive to local variations,
revealing features such as gaps, spirals, and rings
(\citealp{2012ApJ...748L..22M}, \citealp{2018ApJ...863...44A},
\citealp{2020A&A...633A..82G}, \citealp{2022AJ....164..109R},
\citealp{2024A&A...684A..73D}, \citealp{2024A&A...685A..52G},
\citealp{2024A&A...685A..53G},
\citealp{2024A&A...685A..54V}). Instruments such as SPHERE (VLT), GPI
(Gemini), and HiCIAO (Subaru) achieve high angular resolution and
contrast, essential for studying faint scattered light in the presence
of bright central stars. Advances in adaptive optics and coronaraphy
have further enhanced the capability to image disks at high fidelity.

Among scattered light imaging techniques, polarimetric differential imaging (PDI) stands out for its sensitivity to polarised light, which arises from scattering by dust grains. The PDI effectively separates polarised scattered light from unpolarised stellar light, allowing for a detailed characterisation of disk morphology and dust properties (e.g. \citealp{2016A&A...595A.114D, 2016A&A...595A.112G, 2023ApJ...944L..43T}). By using Stokes parameters $Q$ and $U$ to derive polarised flux and polarisation angles, PDI provides robust measurements with minimal self-subtraction artefacts. This makes it particularly suitable for studying disk structures and small-scale features. Other imaging techniques, such as angular differential imaging (ADI, \citealp{2006ApJ...641..556M}) and reference differential imaging (RDI, see e.g. \citealp{2021A&A...648A..26W}), are also used occasionally.

Comparing the Solar System to many of the recently observed
planet-forming disks highlights an intriguing discrepancy: whereas the
Solar System appears to halt around 50 AU (Kuiper Cliff,
\citealp{2004AJ....128.1364B}), many planet-forming disks appear to be
significantly more extended. For some compact disks, the possibility
has been raised that the small extent may be due to recent close
stellar encounters (\citealp{2016A&A...594A..53B};
\citealp{Cuello_2018}). A crucial challenge is to confidently
attribute these cut-offs (truncations) to external encounters and to
distinguish them from inherent factors such as the exponential
decrease in matter density of such disks. To explore this, we turn to
scattered light images of inclined disks, specifically focusing on the
observable backside of these protoplanetary disks. The intention is to
leverage the visibility of the backside to gain insights into the
structural intricacies and matter distribution beyond the disc's outer
boundary. While some well-studied disks appear radially extended,
unbiased ALMA surveys
reveal that most disks are compact, typically <50 AU in size
(\citealp{2019ApJ...882...49L}; \citealp{2025A&A...696A.232G}) and,
thus, smaller than the Solar System.

\begin{figure}
    \centering
    \includegraphics[width=1\linewidth]{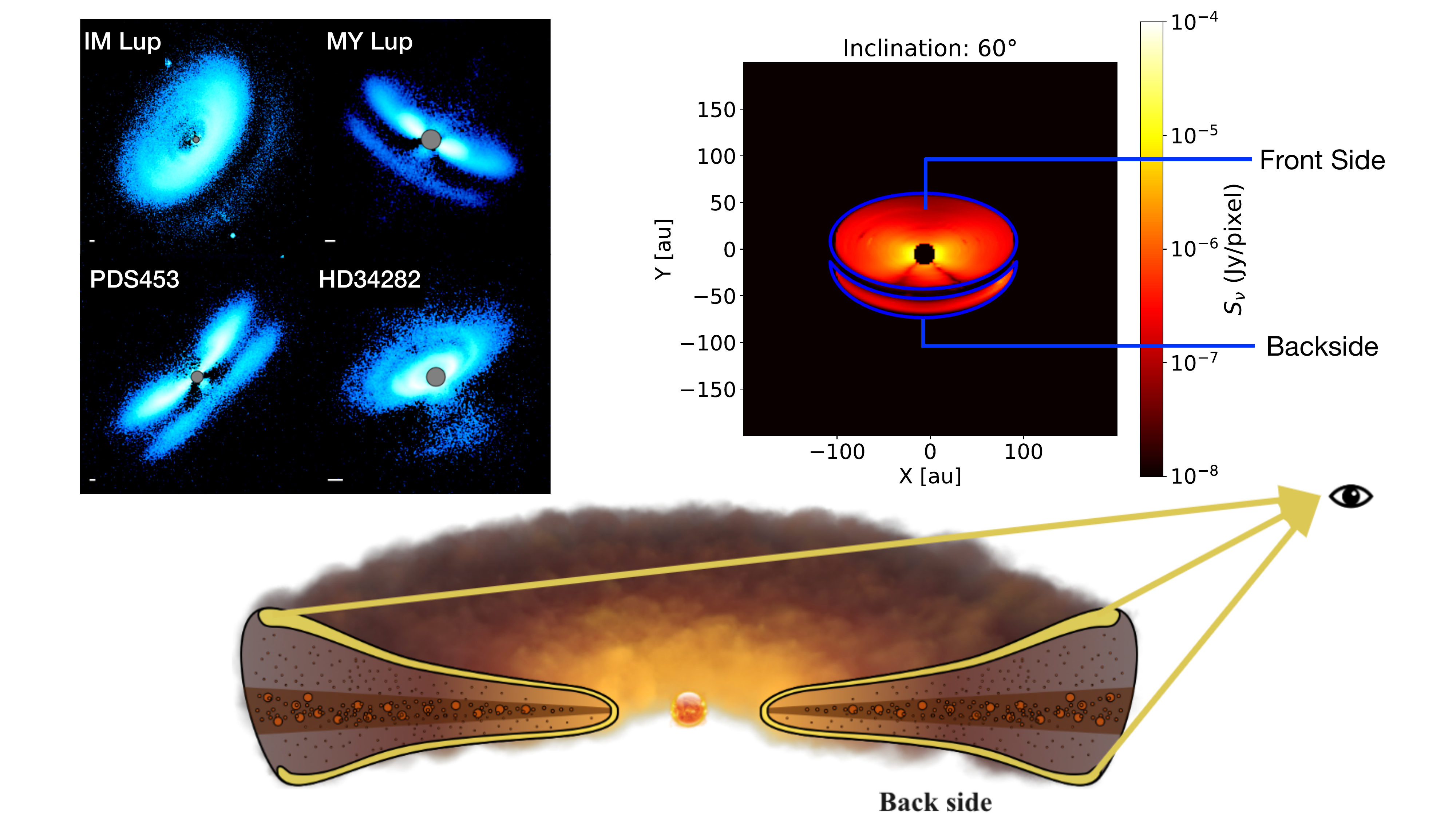}
    \caption{Top left: Examples of disks with observed backside (\citealp{2023ASPC..534..605B}). Top right: Simulated scattered light image of a protoplanetary disk. Bottom: Schematic showing the front and back scattering surfaces of a disk, adapted from the original by Til Birnstiel}
    \label{fig:1}
\end{figure}

Protoplanetary disks have two distinct scattering surfaces – one oriented towards the observer (the top or front side), typically flaring in the observer's direction, and another oriented away from the observer, commonly referred to as the backside, typically flaring in the opposite direction (Figure \ref{fig:1}). In the realm of scattered light imaging, the prominence of the backside becomes apparent when the observer's line of sight intersects with the forward-scattered light emanating from that part of the disk. Figure \ref{fig:1} displays a few protoplanetary disks where the presence of a visible backside has been observed. Investigating the backside of these disks offers us a 3D perspective on the whole disk, including the structure of the disk and properties of the dust that scatters the starlight, since the detectability of the backside hinges on factors such as disk inclination, the scattering properties of dust grains, and the wavelength at which observations are conducted.

Conversely, the absence of discernible backside features in scattered light images suggests potential factors, such as the presence of an optically thick region obscuring the scattered light or an inclination angle minimising backside visibility. Additionally, it could also be attributed to the limited availability of specific dust particles that efficiently scatter light or noise in the observations. Untangling the conditions influencing the observability of the backside significantly contributes to refining our understanding of protoplanetary disk dynamics.

The implications of detecting or not detecting the backside of a disk extend beyond mere observations, as they directly impact the interpretation of data. For instance, in some modelling scenarios, the addition of an extended and optically thick disk component is employed to account for the non-observability of the backside of a luminous disk (e.g. \citealp{2022A&A...668A..25V}). However, this raises the question whether an extended disk is truly necessary to obscure the backside in the observations. In essence, studying the backside of protoplanetary disks and their visibility under specific conditions opens a window into understanding the disk properties better and would aid in constraining disk modelling and refining hypotheses related to disk evolution.

To address this challenge, we employed the radiative transfer code, \texttt{RADMC-3D} (\citealp{2012ascl.soft02015D}), as our primary tool. Conducting a comprehensive parameter study, our objective was to unravel the various factors that influence the visibility of the backside in scattered light, enabling us to later compare simulated results with actual observational data to shed light on why some disks exhibit a visible backside while others, expected to show a visible backside, seemingly do not. This approach involves a synergistic combination of simulations and models, complemented by empirical observations, to enhance our understanding of protoplanetary disks.

To explore the visibility and characteristics of protoplanetary disk backsides, we begin with an overview of the models used in our study, elucidating the vertical and radial structure of disks in section \ref{section_2}. Section \ref{section_3} outlines the methodology for simulating polarised light images using \texttt{RADMC-3D}. In Section \ref{section_4} we present parameter studies to explore how various factors influence backside visibility. The discussion in Section \ref{section_5} addresses observational challenges, including noise and the detectability of backsides at lower inclinations, offering a thorough analysis of our findings. Lastly, section \ref{section_6} summarises our results and their broader implications.

\section{Models}
\label{section_2}

In this section, we define and introduce the radiative transfer model, the structure of the disk, and their underlying physics. Protoplanetary disks consist of a central star surrounded by a mixture of gas and dust that revolve due to the conservation of angular momentum. The equilibrium structure of gas in Protoplanetary disks can be simplified by several key assumptions. The disk mass is negligible compared to the host star, which allows stellar gravity to dominate and neglect disk self-gravity. Efficient radiative cooling leads to a thin disk with a vertically isothermal temperature profile governed by stellar radiation. Radially, the disk achieves hydrostatic equilibrium, with the Keplerian angular velocity (Keplerian orbital frequency) defining the radial structure,
\begin{equation}
\Omega = \sqrt{\frac{GM_\ast}{r^3}}.
\end{equation}
In vertical hydrostatic equilibrium, the density profile follows a Gaussian distribution with
\begin{equation}
\rho_\mathrm{g}(r, z) = \rho_0(r) \exp \left( -\frac{z^2}{2h_\mathrm{p}^2} \right)
\label{equn:gaussian_gas},
\end{equation}
where \(\rho_0(r)\) is the midplane (z = 0) density at radius \(r\), which exhibits the highest gas density.

The scale height and the isothermal sound speed are given by
\begin{equation}
h_\mathrm{p} = \frac{c_\mathrm{s}}{\Omega}
,\\
c_\mathrm{s} = \sqrt{\frac{k_\mathrm{B} T}{\mu m_\mathrm{p}}},
\end{equation}

where \( T \) is the temperature, \( k_\mathrm{B} \) is the Boltzmann constant, \( \mu \) is the mean molecular weight (2.3 for molecular hydrogen), \( m_\mathrm{p} \) = \( m_\mathrm{H} \) is the mass of a proton or hydrogen.

The gas surface density of the disk, \(\Sigma_\mathrm{g}(r)\), is defined as the integral of the volume density \(\rho_\mathrm{g}(r, z)\) over the vertical (\(z\)) direction. Thus, we can write
\begin{equation}
\rho_\mathrm{g}(r, z) = \frac{\Sigma_\mathrm{g}(r)}{\sqrt{2\pi} h_\mathrm{p}(r)} \exp \left( -\frac{z^2}{2h_\mathrm{p}^2} \right).
\end{equation}
Next, we describe the dust distribution. Gas affects solid body dynamics through the drag force it exerts on the dust particles. In the Epstein regime which we are interested in, this force, $\boldsymbol{F}_\mathrm{d}$, takes the simple form
\begin{equation}
\boldsymbol{F}_\mathrm{d} = -\frac{\boldsymbol{v} - \boldsymbol{v}_\mathrm{d}}{\tau_\mathrm{s}},
\end{equation}
where $\boldsymbol{v}$ and $\boldsymbol{v}_\mathrm{d}$ are the gas and dust velocities, respectively. $\tau_\mathrm{s}$ is the dust stopping time. It depends on the dust particle mass density, $\rho_\mathrm{s}$, and its size, $a$, through
\begin{equation}
\tau_\mathrm{s} = \frac{\rho_\mathrm{s} a}{\rho c_\mathrm{s}},
\end{equation}
where $\rho$ is the gas density structure, which follows the Gaussian profile (Equation \ref{equn:gaussian_gas}).

A key dimensionless parameter in the study of dust dynamics within protoplanetary disks is the Stokes number, given by $\text{St} = \Omega_\mathrm{K} \tau_\mathrm{s}$. Using equations 5 and 7,
\begin{equation}
\mathrm{St} = \Omega_\mathrm{K}\tau_\mathrm{s} = \Omega_\mathrm{K}\frac{\sqrt{2\pi} \ h_\mathrm{p} \ \rho_\mathrm{s} \ a}{\Sigma_\mathrm{g}(r)\  c_\mathrm{s}} \exp \left( \frac{z^2}{2h_\mathrm{p}^2} \right),
\end{equation}
which we can evaluate at the midplane $( z = 0 )$ to
\begin{equation}
\mathrm{St_{\mathrm{mid}}} = (\Omega_\mathrm{K}\tau_\mathrm{s})_\mathrm{mid} = \sqrt{2\pi}\frac{ \rho_\mathrm{s}  a}{\Sigma_\mathrm{g}(r)}. 
\end{equation}

The partial differential equation governing the vertical evolution of dust density (\citealp{2004ApJ...614..960S}, \citealp{2004A&A...421.1075D}) is given by
\begin{equation}
\frac{\partial \rho_\mathrm{d}}{\partial t}-\frac{\partial}{\partial z}\left(z \Omega^2 \tau_\mathrm{s} \rho_\mathrm{d}\right)=\frac{\partial}{\partial z}\left[D \rho \frac{\partial}{\partial z}\left(\frac{\rho_\mathrm{d}}{\rho}\right)\right],
\end{equation}
where $\rho_\mathrm{d}$ represents the dust particle density and $D$ denotes the diffusion coefficient capturing turbulent diffusivity. This equation delineates the interplay between vertical settling and turbulent diffusion. In this approximation, we neglect radial migration of grains due to gas drag or other factors (\citealp{1977MNRAS.180...57W}). Additionally, we disregard the accretional evolution of the disk itself or radial mixing phenomena, focusing solely on the vertical motions induced by settling and vertical turbulent stirring.

The time derivative can be disregarded, as we assume a steady-state vertical profile for the density. Upon integrating and rearranging terms, we obtain
\begin{equation}
\frac{\partial}{\partial z}\left(\ln \frac{\rho_\mathrm{d}}{\rho}\right)=-\frac{\Omega^2 \tau_\mathrm{s}}{D} z.
\end{equation}

The simplest solution is to assume that the diffusion coefficient ($D$) is constant, which is a reasonable assumption when the turbulence is homogeneous. Taking that assumption (\citealp{2009A&A...496..597F}), we obtain
\begin{equation}
\rho_{\mathrm{d}}=\rho_{\mathrm{d}, \text { mid }} \exp \left[-\frac{\left(\Omega \tau_{\mathrm{s}}\right)_{\text {mid }}}{\tilde{D}}\left(\exp \left(\frac{Z^2}{2 h^2}\right)-1\right)-\frac{Z^2}{2 h^2}\right],
\end{equation}
where \( \tilde{D} \) is the dimensionless diffusion coefficient defined through \( D = \tilde{D} c_\mathrm{s} h \). The above equation holds on the assumption that the vertical distribution of gas remains Gaussian throughout. Using different grain sizes, we account for the differential settling and turbulent mixing of grains.

We can express \( \tilde{D}=\alpha/\mathrm{Sc} \) in terms of the turbulent $\alpha$ parameter \citep{1973A&A....24..337S} and the Schmidt number Sc \citep{2004A&A...421.1075D}). The Schmidt number has been measured to be of order one when net magnetohydrodynamic (MHD) turbulence is zero (\citealp{2005ApJ...634.1353J}, \citealp{2008A&A...483..815I}).

\begin{figure*}
    \centering
    \includegraphics[width=1\linewidth]{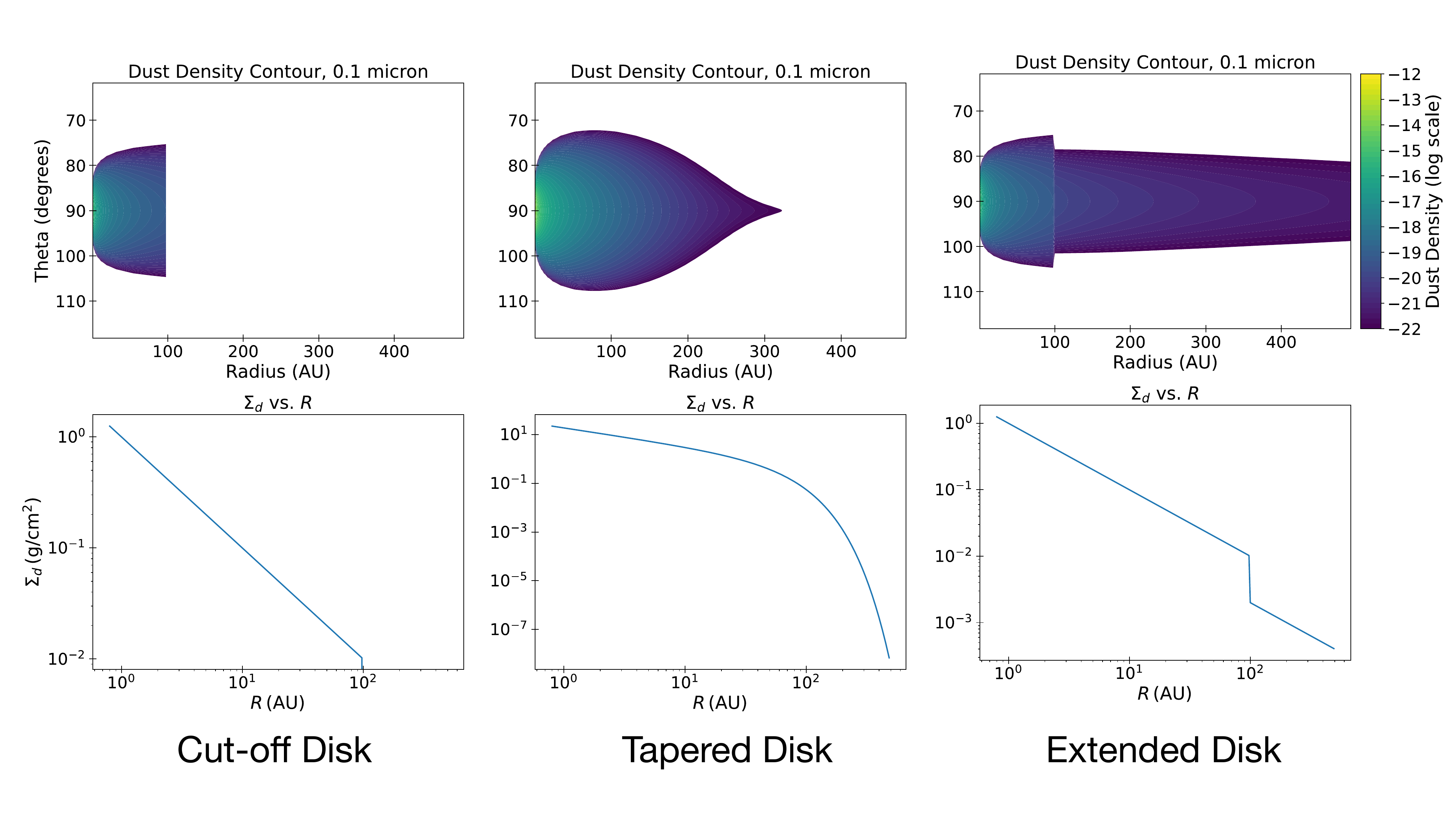}
    \caption[Contour plot of dust density cross-section and Logarithmic plot of surface density.]{Top: Contour plots illustrating the dust density cross-section for a cut-off disk, a tapered disk, and an extended disk, respectively. The horizontal axis represents the radial distance from the central star, while the vertical axis shows the height above and below the disk midplane ($90^\circ$). The colour scale denotes dust density, with warmer colours indicating higher densities. Bottom: Corresponding logarithmic plots of dust surface density as a function of radial distance from the star for each disk type shown above. The vertical axis represents surface density, and the horizontal axis denotes the distance from the star.}
    \label{fig:3}
\end{figure*}

Radial drift can significantly deplete large grains (e.g. $>100\,\mu\mathrm{m}$) in the outer regions of protoplanetary disks. However, the scattering signal in the near-infrared (e.g. H-band, $1.6\,\mu\mathrm{m}$) is dominated by much smaller grains. Grains smaller than $\sim1\,\mu\mathrm{m}$ typically contribute more than 50\% of the opacity in the H band, and grains smaller than $\sim10\,\mu\mathrm{m}$ contribute over 80\%. The small grains are the ones that do the scattering in the upper disk layers because they do not settle much, and they are also responsible for the optical depth through the disk that may obscure radiation from the backside. These small particles are well coupled to the gas due to their low Stokes numbers, resulting in radial velocities nearly identical to the gas and minimal radial drift over disk lifetimes. Observationally, this is supported by systems such as IM Lup and TW Hya (\citealp{2020ARA&A..58..483A}, \citealp{2023ASPC..534..501M}, \citealp{2024ARA&A..62..157B}), where the scattered-light disk is far more extended than the millimetre continuum.

For the studies in this paper, we consider three different parametrisations for the mass distribution in the outer disk: cut-off disks, tapered disks, and extended disks. These different dust distribution models allow us to explore how various outer disk configurations affect the resulting scattered light images and enhance our understanding of the disc's backside visibility.

\subsection{Cut-off disks}
\label{sec:2.3}

Cut-off disks are characterised by a sharp outer edge where the dust density abruptly drops to zero. We assume that the surface density $\Sigma(R)$ follows a negative power law with an exponent $\gamma$ as given by
\begin{equation}
\Sigma(R) = 
\begin{cases}
\Sigma_0 \left(\dfrac{R}{R_\mathrm{c}}\right)^{-\gamma}, & \text{for } 0.8\ \mathrm{AU} \leq R \leq R_{\text{out}} \\
0, & \text{otherwise}
\end{cases}
\label{eq:cutoff-disk}.
\end{equation}

For simplicity, we set $R_\mathrm{c} = 1$ AU. This parametrisation offers the simplest form for the disk surface density, effectively mimicking the sharp inner disk truncation that can be caused by disk photoevaporation or by dynamical interactions with a giant planet or a flyby/companion star. However, it has the drawback of introducing an unrealistic sharp edge in $\Sigma$ at the disc's outer radius. Dust density contour and surface density versus distance from the star of an example cut-off model with $R_{\text{out}}=100$ AU and $\gamma =  1.0$ is given in Figure \ref{fig:3}.

\subsection{Tapered disks}
\label{sec:2.4}

In tapered disks, the dust density gradually decreases towards the
outer edge following an exponential or other tapering function. We
chose the tapered-edge axisymmetric disk model given by
\citealp{1974MNRAS.168..603L} for the disk surface density of a
viscous Keplerian disk (\citealp{1998ApJ...495..385H}). In this
scenario, the disk undergoes viscous spreading, with mass steadily
accreted inwards while some outer disk material is redistributed
further outwards to conserve angular momentum. The resulting surface
density profile is given by
\begin{equation}
\Sigma(R)=\Sigma_c\left(\frac{R}{R_\mathrm{c}}\right)^{-\gamma} \exp \left[\left(-\frac{R}{R_\mathrm{c}}\right)^{2-\gamma}\right] \quad (R > 0.8\,\mathrm{AU})
\label{eq:2.28}.
\end{equation}

Here, $\Sigma_c$ is the surface density at the characteristic radius $R_\mathrm{c}$, and $\gamma$ is the power-law index. According to this equation, the surface density ($\Sigma$) follows an exponential taper outside the critical radius $R_\mathrm{c}$. This tapering function creates a realistic disk profile with a smooth density decrease and gradual extension into space, providing a physically motivated alternative to the cut-off model's abrupt truncation. Hence, the model is quite often used in literature to characterise both normal and transition disks (e.g. \citealp{2009ApJ...700.1502A}, \citealp{2017ApJ...851...56W}). Dust density contour and surface density versus distance from the star of an example tapered disk model with $\gamma =  0.75$ and $R_\mathrm{c} = 50 $ AU is given in Figure \ref{fig:3}.

\subsection{Extended disk}
\label{sec:2.5}

We use the term ‘extended disk’ to describe a setup where a cut-off disk is surrounded by additional material with a significantly lower surface density compared to the profile of the main disk. Figure \ref{fig:3} illustrates an example of an extended disk model, showing both the dust density contour and the surface density as a function of distance from the star. We take the inner radius of 0.8 AU, similar to the previous cases. In this model, the inner disk (0.8 to 100 AU) adheres to a specific surface density profile—specifically, the cut-off disk profile discussed in Section \ref{sec:2.3}, while the outer disk (100 to 500 AU) follows a modified surface density profile. In this case, the outer disc's surface density is reduced by a factor of 5 from the inner disc's cut-off profile.

Extended disk models could be used as an experimental setup that provides full control over the outer surface density to adjust disk transparency, helping us better understand how much matter must be present in the outer disk to effectively block the backside. Adding an additional extended outer disk is occasionally used as a method to hide the backside in PPD models, when the actual observational data do not show a backside (e.g. \citealp{2022A&A...668A..25V}).

\subsection{Disk illumination and heating}
To observe these protoplanetary disks, we relied on the illumination and heating provided by their host stars, which enable the detection of scattered light from the disc's surface layers. Analysing this scattered light requires an understanding of the disc's temperature distribution, a fundamental parameter in characterising its physical and thermal properties. Most disks are recognised to exhibit a flaring geometry, where the aspect ratio ($ h_\mathrm{p} / r$) increases with radial distance from the star, enabling the outer regions to intercept more stellar photons and sustain higher temperatures at larger radii. \cite{Chiang_1997} proposed a two-layer passive disk model with a hot, irradiated surface layer that re-radiates half the stellar flux and a cooler, vertically isothermal interior that absorbs and re-emits the remainder as thermal radiation. This model, combined with the flaring geometry, successfully reproduces observed spectral energy distributions and accounts for enhanced mid- to far-infrared emission (\citealp{2001ApJ...560..957D}). The irradiative flux at a distance \( r \) from the star is given by
\begin{equation}
F_{\text{irr}} = \alpha_{\text{irr}} \frac{L_\star}{4 \pi r^2},
\end{equation}
where \( L_\star \), the stellar luminosity, is calculated using the Stefan-Boltzmann law, \( L_\star = 4 \pi R_\star^2 \sigma T_\star^4 \). The flaring irradiation incidence angle \( \alpha_{\text{irr}} \), which is the angle between the stellar radiation and the local disk surface, is assumed constant and fixed at 0.05 radians in our models. Assuming thermal equilibrium, the disk temperature equals the dust midplane temperature as in
\begin{equation}
T_{\text{disk}} = \left( \frac{F_{\text{irr}}}{\sigma} \right)^\frac{1}{4},
\end{equation}
where \( \sigma \) is the Stefan-Boltzmann constant.

\section{Simulating disk images}
\label{section_3}

Disks contain a range of dust species, from small particles to large grains, with their distribution influenced by size: smaller grains mix well with the gas, while larger grains settle towards the midplane. In this section, we explore the methodology for simulating polarised light images using \texttt{RADMC-3D} \citep{RADMC3D} since contemporary scattered light observations of disks predominantly utilise polarised light (e.g. \citealp{2018ApJ...863...44A}, \citealp{2020A&A...633A..82G}, \citealp{2024A&A...685A..54V}, \citealp{2024A&A...685A..53G}, \citealp{2024A&A...685A..52G}), facilitated by PDI.

\subsection{Scattering}
\label{sec:3.1.2}

All astronomical sources are partially polarised to some degree, and the measurement of polarisation is described using the Stokes-Mueller formalism (\citealp{1851TCaPS...9..399S}, \citealp{1960ratr.book.....C}). The Stokes vector is defined as
\begin{equation}
{S} = \begin{pmatrix}
I \\
Q \\
U \\
V
\end{pmatrix}. 
\end{equation}
\(I\) denotes the intensity regardless of polarisation. \(Q\) and \(U\) describe the states of linear polarisation, and \(V\) represents circular polarisation.

Interactions between matter and polarised radiation are mathematically described using \(4 \times 4\) Mueller matrices. These matrices operate on the Stokes vector of incoming radiation, \(\mathbf{F}_{\text{in}}\), to yield the scattered Stokes vector, \(\mathbf{F}_{\text{out}}\), which describes the polarisation state of light after interaction.

For spherical or randomly oriented dust grains without any preferential helicity, the scattering matrix simplifies due to symmetry. The outgoing Stokes vector \(\mathbf{F}_{\text{out}}\) is given by
\begin{equation}
F_{\text{out}} = \begin{pmatrix}
F_{I, \text{out}} \\
F_{Q, \text{out}} \\
F_{U, \text{out}} \\
F_{V, \text{out}}
\end{pmatrix} = \frac{m_{\text{grain}}}{r^2} \begin{pmatrix}
Z_{11} & Z_{12} & 0 & 0 \\
Z_{12} & Z_{22} & 0 & 0 \\
0 & 0 & Z_{33} & Z_{34} \\
0 & 0 & -Z_{34} & Z_{44}
\end{pmatrix} \begin{pmatrix}
F_{I, \text{in}} \\
F_{Q, \text{in}} \\
F_{U, \text{in}} \\
F_{V, \text{in}}
\end{pmatrix}.
\end{equation}

For randomly oriented non-helical particles, the scattering opacity, \(\kappa_{\text{scat}}\), can be calculated as
\begin{equation}
\kappa_{\text{scat}} = \oint Z_{11} \, d\Omega = 2\pi \int_{-1}^{+1} Z_{11}(\mu) \, d\mu.
\end{equation}

Since we consider anisotropic scattering, the angular distribution of scattered radiation is described by the scattering phase function (\(\Phi\)), normalised over all directions to unity. It can be calculated directly from the scattering matrix via
\begin{equation}
\Phi(\mu)  = \frac{4\pi}{\kappa_{\mathrm{scat}}}\,Z_{11}(\theta).
\end{equation}

\subsection{Opacities in the models}
\label{sec:4.1.3}

Opacities are fundamental for any model using {\texttt{RADMC-3D}}, and we must provide opacity files that contain the dust particle opacities necessary for the simulation. For grains of different sizes, we must consider their size distribution. We adhere to the MRN (Mathis, Rumpl, Nordsieck) size distribution (\citealp{1977ApJ...217..425M}):
\begin{equation}
n(a) \, da \propto a^{-3.5} \, da,
\end{equation}
where $a$ is the radius of the grain and $n(a) \, da$ is the number of grains with sizes between $a$ and $a+da$ per unit volume. We define this power law to extend from $a_{\mathrm{min}}$ to $a_{\mathrm{max}}$. The total dust density, $\rho$, is then given by
\begin{equation}
\rho = \int_{a_{\mathrm{min}}}^{a_{\mathrm{max}}} m(a) \, n(a) \, da, \\ 
m(a) = \rho_\mathrm{s} \frac{4\pi}{3}a^3,
\end{equation}
where $m(a)$ is the mass of grain, $\rho_\mathrm{s}$ is the material density of the grain material, typically ranging between 1 and 3.6 g/cm$^3$, depending on the material.

Opacity files are generated for 15 distinct grain sizes ranging from 0.05 to 2000 microns (0.05, 0.1, 0.2, 0.5, 1.0, 2.0, 5.0, 10.0, 20.0, 50.0, 100.0, 200.0, 500.0, 1000.0, 2000.0 $\mu$m), with each grain size treated as an independent dust species with a unique spatial density distribution. This approach facilitates realistic simulations by accurately representing the effects of turbulent mixing and settling, such as the concentration of larger grains near the midplane and the broader dispersion of smaller grains. The opacity files, tailored to each grain size, were created using \texttt{optool} (\citealp{2021ascl.soft04010D}), a specialised tool for calculating dust opacities and scattering matrices.
 
We used the DIANA standard opacities (\citealp{2016A&A...586A.103W}) to provide an accurate and realistic representation of dust in protoplanetary disks. These opacities are based on standard assumptions for disk models, featuring a specific amorphous pyroxene (\(\text{Mg}_{0.7}\text{Fe}_{0.3}\text{SiO}_3\)) and carbon, in a mass ratio of 0.87 : 0.13, and a porosity of 25\%. The pyroxene (70\% Mg) has a material density of 3.01 g/cm$^3$, while the carbon (C-z) has a material density of 1.80 g/cm$^3$. The resulting grain mixture in vacuum has a material density of 2.08 g/cm$^3$.

\begin{figure*}
    \centering
    \includegraphics[width=1\linewidth]{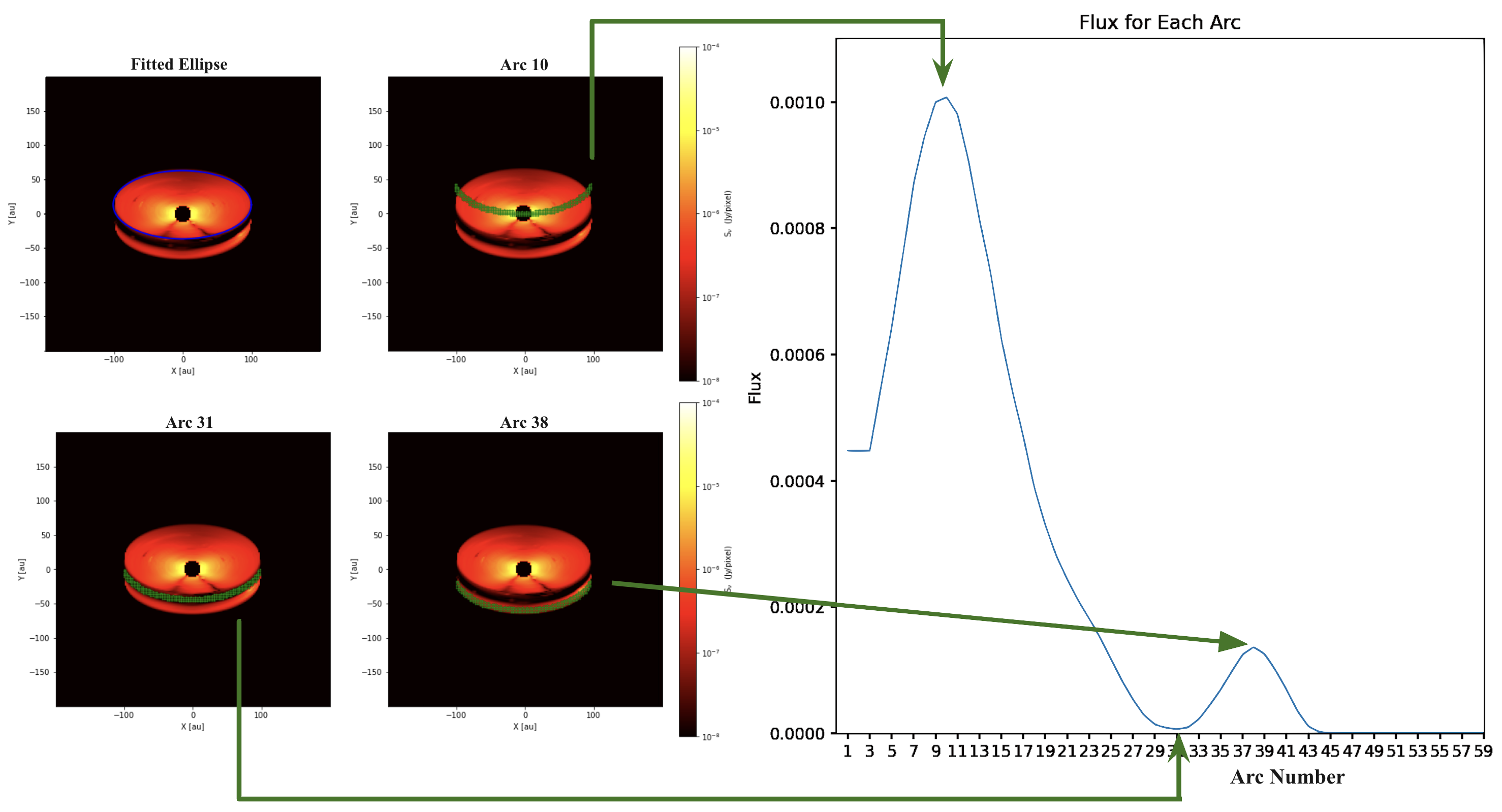}
    \caption[Illustration of measuring the backside using integrated flux in a ‘crescent’.]{Illustration of measuring the backside using integrated flux in a ‘crescent’. The left-hand side shows the disk image with a fitted ellipse at the top and three instances of a solid crescent moving through the image. The right-hand side displays the plot of integrated flux versus the index of the crescent's position. Green arrows indicate points on the plot corresponding to the instances shown on the left-hand side.}
    \label{fig:4.2}
\end{figure*}

\subsection{Imaging}
\label{sec:4.1.4}

Images were created after determining the dust temperature using the thermal Monte Carlo run. The number of pixels in the image, the image size, and the inclination and rotation relative to the observer was input. The observer was situated at a distance of 140 parsecs (pc), comparable to the distances of local star-forming molecular clouds such as Taurus-Auriga, Ophiuchus, Aquila S, Lupus, and Chameleon ($\sim$ 120-200 pc, Gaia Collaboration \citealp{2018}, \citealp{2020A&A...633A..51Z}). The image is square, spanning 400 au (-200 to 200) or 2.86 arcseconds along both the $x$ and $y$ axes, and is composed of a grid of 200 $\times$ 200 pixels, with a pixel scaling of 14.28 milliarcsecond (mas) per pixel. The colour bar was set to display polarised intensity (PI, Flux density $S_\nu$ in jansky/pixel) in a log scale ranging from $10^{-8}$ to $10^{-4}$, with $10^{-4}$ representing the brightest intensity.

The polarised intensity was obtained by combining the Stokes Q and U components in the following equation: $PI = \sqrt{Q^2 + U^2}$. In theory, commonly in actual observations, $Q_{\phi}$ (signal) and $U_{\phi}$ (no signal/noise) are produced, but in practice, $Q_{\phi}$ and PI are nearly identical (\citealp{benisty2022optical}). When the inclination is $0^\circ$, we see the disk face-on, and when it is $90^\circ$, we see it edge-on. The rotation did not affect the images we produced, as the disk was considered to be axis-symmetric.

Unless otherwise noted, all the images were produced in H band (1.65 $\mu$m) and a coronagraph was also added to the images to make our images resemble actual observations more closely. We used a coronagraph resembling the Lyot coronagraph \citep{2011ExA....30...39C, 2011ExA....30...59G}, which is normally used in H-band SPHERE observations. It has a radius of 92.5 mas (12.95 au in our images), which falls well within the typical coronagraph radii in observations (typically at 10–15 au, \citealp{2015A&A...578L...6B, 2016A&A...595A.112G}). The top right panel in figure \ref{fig:1} is an example of the resulting images.

\subsection{Measuring the backside}
\label{sec:4.1.5}

To quantify the observability of the backside, a method was devised to measure the integrated flux in that region. An illustration of this method is provided in Figure \ref{fig:4.2}. An ellipse was fitted onto the front side of the disk, and since the backside typically mirrors the shape of the outer edges of the front side, the edge of this ellipse was extrapolated to create a solid arc referred to as a ‘crescent’ (see the left-hand side of Figure \ref{fig:4.2}). This crescent was then systematically moved down through the disk image, integrating the flux within its pixels at each position. The resulting integrated flux versus index or position of the crescent plot, as depicted on the right-hand side of Figure \ref{fig:4.2}, shows distinct features: a primary peak (Arc 10) corresponding to the crescent's alignment with the central star; a local minimum (Arc 31) representing the dark midplane between the front and backsides; and a smaller peak (Arc 38) indicative of the integrated flux from the backside as the crescent intersects it. The thickness of the crescent we created is set to 4 pixels (8 au), corresponding to the angular resolution of the VLT-SPHERE telescope ($8.2$ metres, calculated as the angular resolution ($\theta) = 1.22 \times \lambda / D$).

\section{Parameter studies}
\label{section_4}

In this section, we analyse the impact of various parameters on the
visibility of the backside of a disk in scattered light imaging,
focusing on different radial distributions and their corresponding
parameters. General details and recurring properties of the images are
explained in Section \ref{sec:4.1.4}. By default, the central star was
assumed to have a mass of \( M_\star = 2.4\,M_\odot \), a radius of
\( R_\star = 2.4\,R_\odot \), and an effective temperature of
\( T_\star = 10{,}000\,\text{K} \). This ensures that the star is
sufficiently luminous, making the visibility of the backside
predominantly dependent on the disc's properties. We restrict our
study here to this case, which is designed for an intermediate-mass
young star. Some of the observational data also pertain to lower-mass T
Tauri stars. However, the visibility of the backside is related to the
stellar luminosity only insofar as the overall S/N of the
image depends on it, but not the relative visibility of front and
backsides. Since we do not study colour effects in the image, this
finding is sufficient for the present study. We examined each
parameter and create images at various inclinations, ranging from
\( 90^\circ \) (edge-on) to \( 30^\circ \), to understand the impact
of inclination with respect to the observer on the resulting
image. Alongside the images of disks depicting each parameter and
inclination, an annotated heat map was included. The colour and value
in each cell of the heat map indicate the integrated flux from the
disc's backside under specific parameter conditions.

\subsection{Cut-off disks}
\label{sec:4.2.1}

\begin{figure*}
    \centering
    \includegraphics[width=\textwidth]{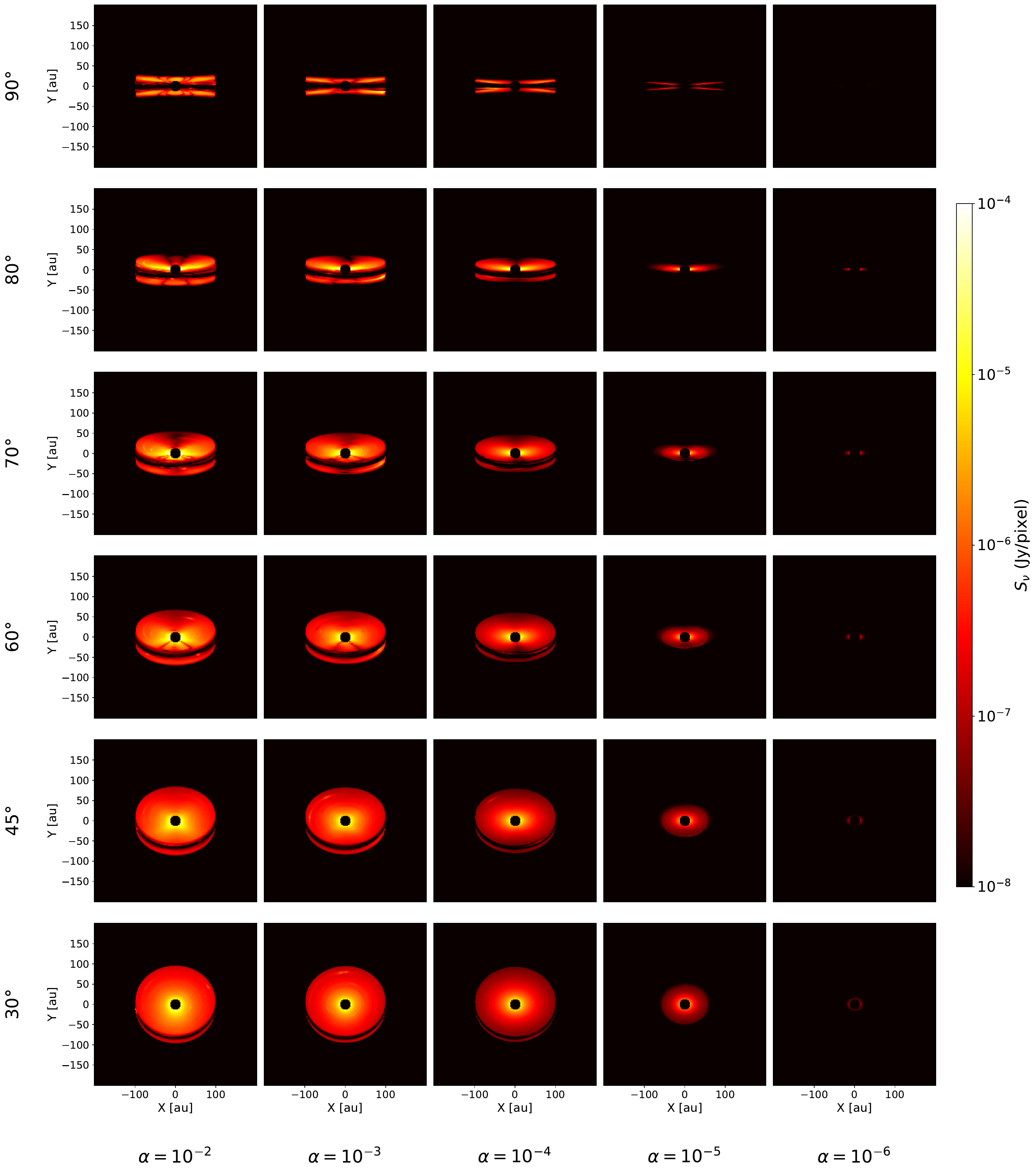}
    \caption{Parameter study of cut-off disks with $\Sigma_\mathrm{d}$ = 1 g/cm$^2$ at $R_\mathrm{c} = 1$ AU for varying the Shakura-Sunyaev turbulence parameter ($\alpha$). Each column represents observations of the disk at different inclinations for a specific $\alpha$ value, with corresponding $\alpha$ values listed on the x-axis. Note that the turbulence parameter $\alpha$ decreases from left to right, indicating reduced turbulence in the images towards the right.}
    \label{fig:PS_cut-off}
\end{figure*}

\begin{figure*}
    \centering
    \includegraphics[width=1.0\linewidth]{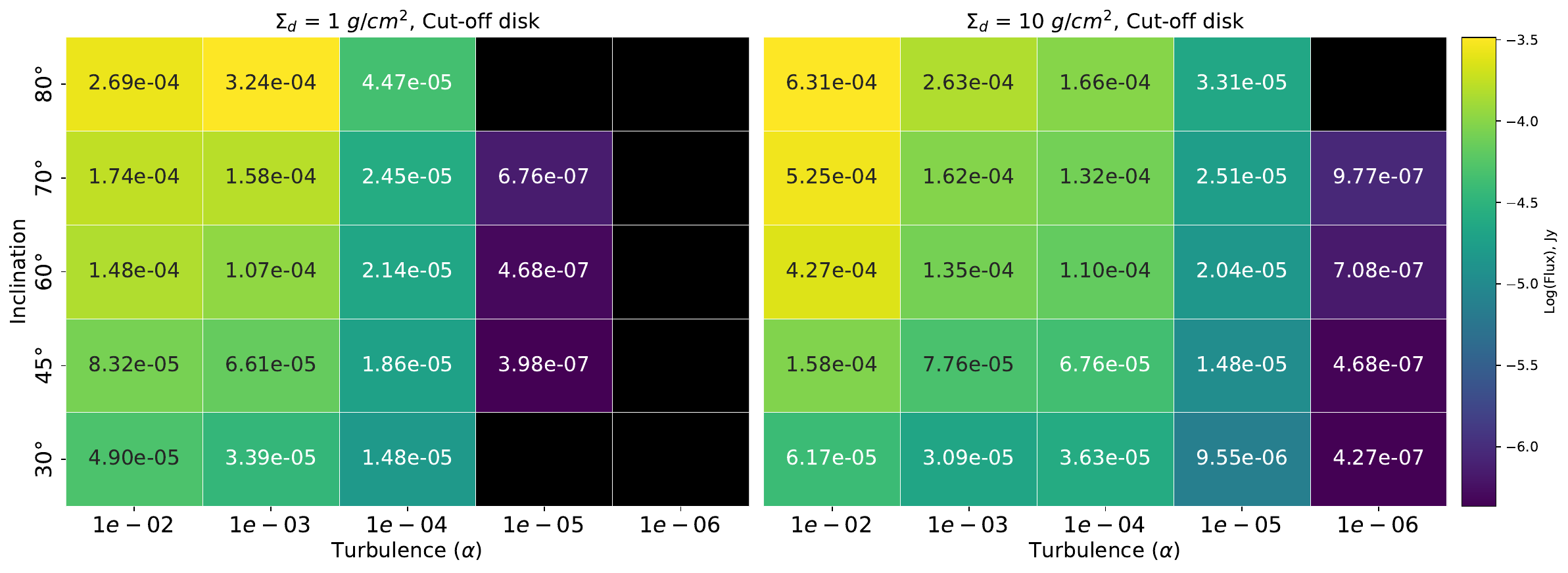}
    \caption[Annotated heat map showing integrated backside flux in
    parameter studies of cut-off disks.]{Annotated heat map showing
      integrated backside flux in parameter studies of cut-off
      disks. The left panel shows the flux for a surface density
      $\Sigma_0$ (dust) fixed at 1 g/cm$^2$ at $R_\mathrm{c} = 1$ AU,
      while the right panel shows the flux for 10 g/cm$^2$ at
      $R_\mathrm{c} = 1$ AU. The cells in brighter yellow
      indicate higher flux, and the cells in darker blue indicate
      lower flux, with values in jansky. The x-axis represents the
      Shakura-Sunyaev turbulence parameter ($\alpha$), and the y-axis
      represents the inclination angles. Black cells represent disks
      where the backside could not be detected (absence of secondary
      peak, refer Figure \ref{fig:4.2})}
    \label{fig:cut_off_heatmap}
\end{figure*}

We examined the visibility of the backside in cut-off disks (Section \ref{sec:2.3}) by changing parameters. Figure \ref{fig:PS_cut-off} presents parameter studies of cut-off disks, which are truncated $(R_{\text{out}})$ at 100 AU from the centre (star), with a surface density $\Sigma_0$ (dust) fixed at 1 g/cm$^2$ and 10 g/cm$^2$ at $R_\mathrm{c} = 1$ AU, respectively. The analysis spans inclinations ranging from $90^\circ$ to $30^\circ$. The Shakura-Sunyaev turbulence parameter ($\alpha$) is varied from $10^{-2}$ to $10^{-6}$, making the disk more settled at lower values of $\alpha$. The surface density follows a negative power law (Equation \ref{eq:cutoff-disk}) for these cut-off disks, with a power-law factor $\gamma = 1.0$ maintained across all images.

Figure \ref{fig:cut_off_heatmap} depicts an annotated heat map illustrating the integrated backside flux for the two parameter studies. Measurements exclude the 90-degree edge-on orientation where the two sides exhibit similar brightness due to symmetry. Generally, both sides of edge-on disks are visible in observations. A discernible trend shows that disks with higher surface density appear brighter in disk images, correlating with increased backside flux. This phenomenon likely arises because denser disks contain more dust, thereby increasing the scattering of light towards the observer. Similarly, as the disc's inclination decreases, the integrated flux also diminishes. This is primarily due to geometry, whereby the backside of the disk becomes thinner with decreasing inclination, resulting in reduced integrated flux. Moreover, decreasing the turbulence $\alpha$ parameter results in a more settled disk appearance and decreased brightness, making it more challenging to detect the disc's backside and resulting in lower integrated flux.

\subsection{Tapered disks}
\label{sec:4.2.2}

\begin{figure*}
    \centering
    \includegraphics[width=1\linewidth]{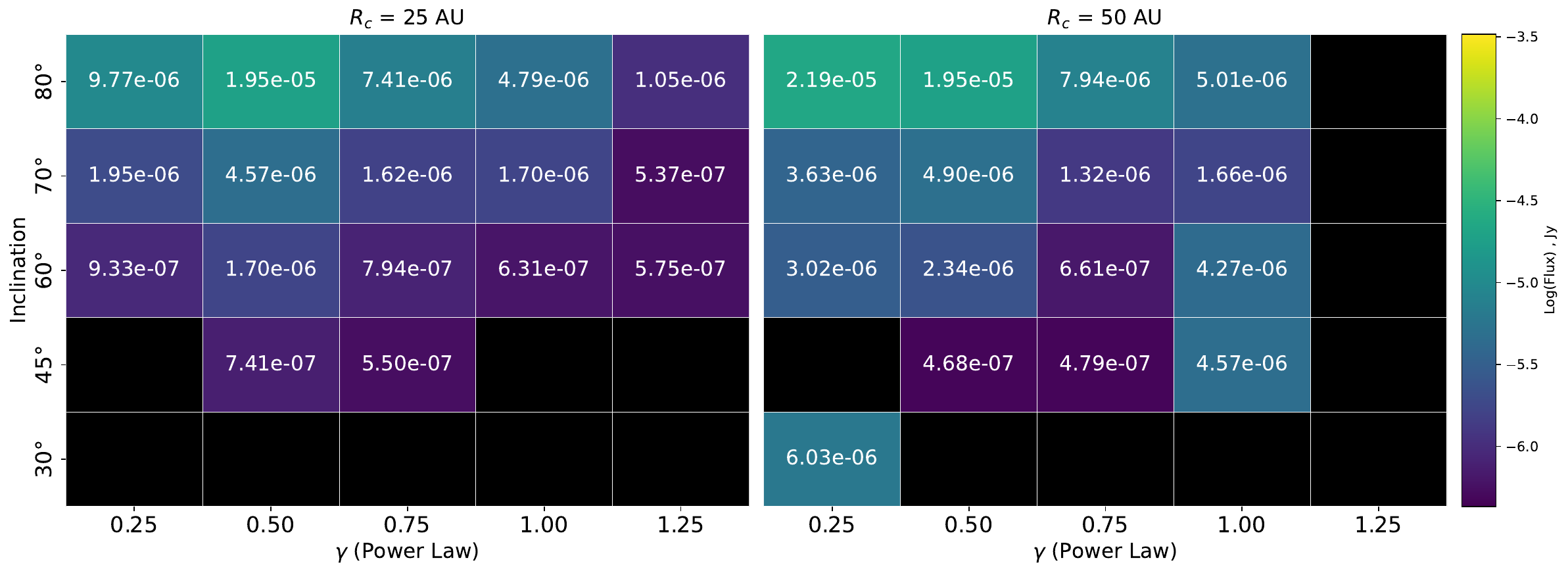}
    \caption[Annotated heat map showing integrated backside flux in
    parameter studies of tapered disks.]{Annotated heat map showing
      integrated backside flux in parameter studies of tapered
      disks. The left panel shows the flux for disks with a
      characteristic radius, $R_\mathrm{c} = 25$ AU, while the right
      panel shows the flux for disks with $R_\mathrm{c} = 50$
      AU. Brighter yellow cells indicate higher flux, and darker blue
      cells indicate lower flux, with values in jansky. The x-axis
      represents the negative power law ($\gamma$), and the y-axis
      represents the inclination angles. Black cells represent disks
      where the backside could not be detected (absence of secondary
      peak, refer Figure \ref{fig:4.2}) }
    \label{fig:4.8}
\end{figure*}
In tapered disks, as detailed in section \ref{sec:2.4}, the dust density gradually decreases outwards following an exponential function. This model is more realistic compared to cut-off disks because it incorporates viscous spreading, as described by Equation \ref{eq:2.28}. Here, mass migrates inwards due to accretion, while some outer disk material redistributes outwards to maintain angular momentum conservation.

Figure \ref{fig:combined_tapered} illustrates the parameter study of tapered disks, examining the effects of increasing the exponent of the negative power law ($\gamma$) as described in equation \ref{eq:2.28} while keeping the characteristic radius ($R_\mathrm{c}$) fixed at 25 AU and 50 AU, respectively. The surface density ($\Sigma_c$, dust) at the characteristic radius is constant at 1 g/cm$^2$, and the turbulence \(\alpha\) parameter is set to 0.001.

The front side of the disk remains visible across nearly all inclinations and \(\gamma\) values, but the backsides are challenging to detect visually. Comparing the images and the annotated heat map in Figure \ref{fig:4.8} with the annotated heat map of cut-off disks in Figure \ref{fig:cut_off_heatmap}, it is evident that the integrated flux values have significantly declined for tapered disks. As discussed in the previous chapter, this reduction in flux can be attributed to the presence of sufficient dust in the outer disk. The exponential decay of surface density ensures enough dust to attenuate the scattered light from the backside but not enough to scatter it in our direction. Similar to the cut-off disks, a decrease in inclination generally results in lower integrated flux values for tapered disks. The black cells in the heat map indicate disks where a secondary maximum could not be detected, and thus no true backside could be detected. 

\subsection{Extended disks}
\label{sec:4.2.3}

To validate the hypothesis that sufficient dust in the outer disk can block the visibility of the backside, we simulated an experimental setup similar to that described in the previous chapter. We set up an experimental disk configuration as detailed in Section \ref{sec:2.5}, with an inner disk extending up to 100 AU following a specific surface density profile, and an outer disk extending from 100 AU to 500 AU with a modified surface density.

In this setup, we began with a cut-off disk that terminates abruptly at 100 AU, which typically reveals the backside as seen in the parameter studies in Section \ref{sec:4.2.1}, with the turbulence parameter \(\alpha\) set to 0.001. We then introduced an extended outer disk ranging from 100 AU to 500 AU, with a surface density reduced by a specific factor compared to the inner disk. The outer disk contains the same range of dust species (all 15, from small to large) as the inner disk. The surface density for the cut-off disk follows a negative power law (Equation \ref{eq:cutoff-disk}), with the power law exponent \(\gamma\) fixed at 1.0 and \(\Sigma_0\) (dust) fixed at 1 g/cm\(^2\) at \(R_\mathrm{c}\) (1 AU). The results of the parameter study are shown in Figure \ref{fig:4.9}.

The parameter study reveals that when the outer disk has sufficient mass, for instance, when the surface density \(\Sigma_\mathrm{d}\) is halved, the outer disk effectively obscures the backside. Conversely, when the surface density of the outer disk is reduced further, the outer disk becomes progressively less optically thick, making the backside increasingly visible.

\section{Discussion and analysis}
\label{section_5}

At the time of writing this paper, approximately 200 disks were observed in scattered light, yielding high-resolution images. Out of these, only around 18 disks show evidence of backside detection, constituting less than 10\% of the total. Specifically, the disks where backside detection has been observed include PDS 111, IM Lup, MY Lup, DoAr 25, PDS 453, V599 Ori, V1012 Ori, PDS 144, J1608, BN CrA, J1609, T CrA, HD34282, DG Tau, GM Aur (faint), V409 Tau (faint), Haro 1-1, and V721 CrA (most but not all of these disks are published; \citealp{2018ApJ...863...44A}, \citealp{2019A&A...624A...7V}, \citealp{2020A&A...633A..82G}, \citealp{2021A&A...649A..25D}, \citealp{2023A&A...671A..82R}, \citealp{2024A&A...685A..54V}, \citealp{2024A&A...685A..53G}, \citealp{2024A&A...688A.149D}, Columba et al. in prep, Garufi et al. in prep.).

Among these, seven disks exhibit very high inclinations or are nearly edge-on: MY Lup, DOAr 25, PDS 453, V1012 Ori, PDS 144, T CrA, and J1608. Notably, HD34282, V599 Ori, and V721 CrA are also highly inclined, though not to the extent of being edge-on, likely having inclinations around \( 60^\circ-70^\circ\).

These observations underscore the rarity of backside detection in disks. Our simulations and parameter studies of more physically realistic tapered disks corroborate these findings, showing that most disks visually obscure their backside. Instances where backside detection occurs typically involve very high inclinations or nearly edge-on disks.

This raises two important questions: Primarily, how does noise in actual observations influence our ability to detect backside features? Secondly, why do we occasionally observe disks with detectable backsides at lower inclinations?

\begin{figure*}
    \centering
    \includegraphics[width=1\linewidth]{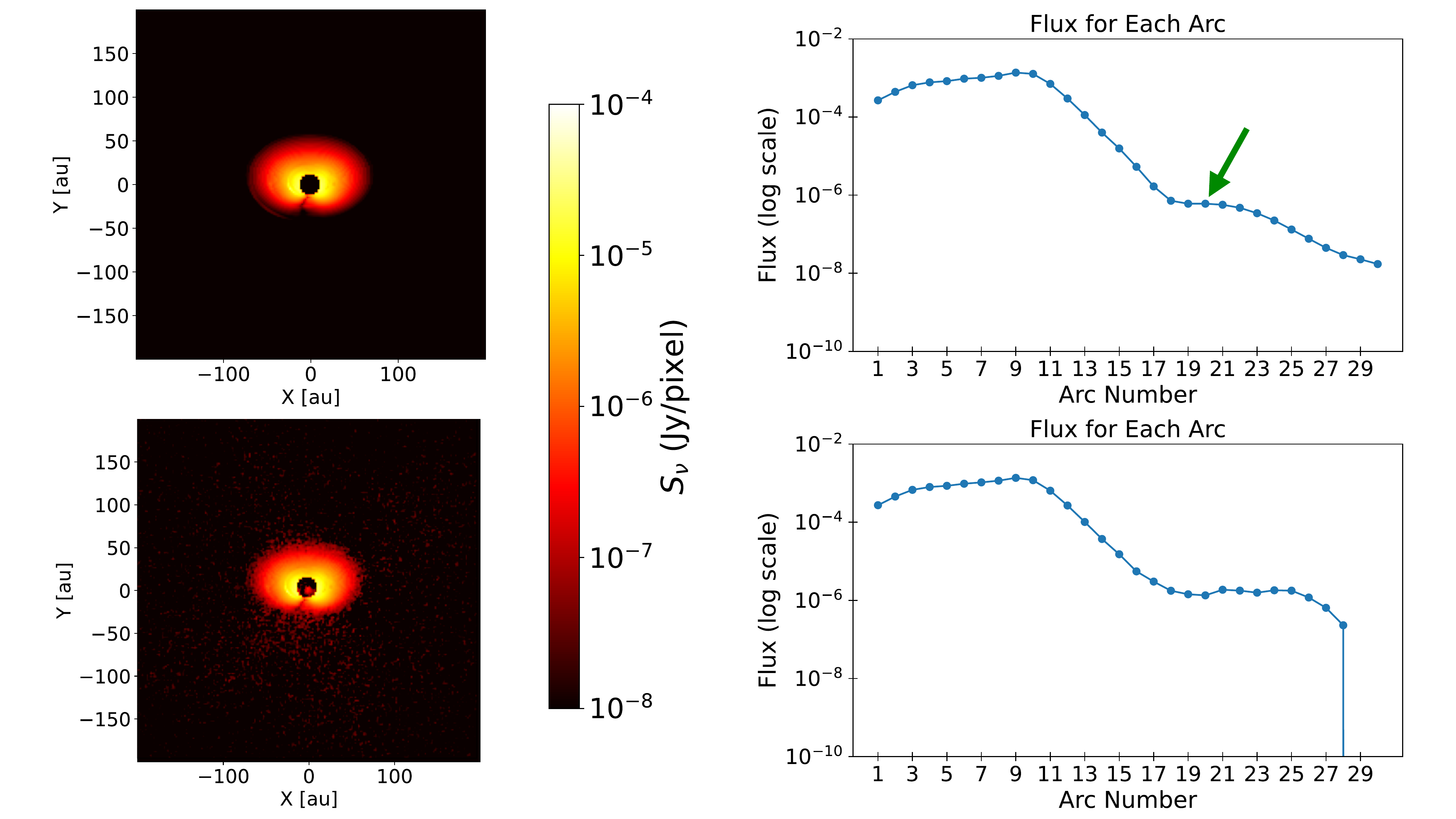}
    \caption[Addition of noise to a simulated disk image.]{Impact of noise addition on a simulated disk image. The top row presents an image of a tapered disk with $R_\mathrm{c} = 25\,\text{au}$ and a power-law exponent $\gamma = 0.75$ at an inclination of $45^\circ$. Correspondingly, the right panel illustrates the integrated flux as the ‘crescent’ (refer to Figure \ref{fig:4.2}) traverses vertically through the disk image on the left, plotted on a logarithmic scale for integrated flux on the y-axis. The green arrow highlights the position of a minor secondary peak representing the backside flux. In the bottom row, the left panel displays the same simulated disk image with added $U_{\phi}$ noise, while the right panel shows the integrated flux as the ‘crescent’ moves down through the noise-added image, also plotted on a logarithmic scale for integrated flux on the y-axis.} 
    \label{fig:5.1}
\end{figure*}

\subsection{Noise in observations}
\label{sec:5.1}

Polarimetric observations of disks using the PDI technique yield two files: \(Q_{\phi}\) and \(U_{\phi}\). Theoretically, all the signal should be in \(Q_{\phi}\), with \(U_{\phi}\) containing no signal. However, in practice, \(U_{\phi}\) does contain some signal, which can be considered systematic noise in the observations.

We utilised the \(U_{\phi}\) file from an actual observation of IM LUP to introduce noise into our model images. This is demonstrated on the bottom row of Figure \ref{fig:5.1}. When the backside signal is not significant enough, the noise can obscure it, preventing a secondary peak (representing the backside) from appearing on our integrated flux plot as we move down our flux integrating ‘crescent’, as shown on the right-hand side of the same figure. This implies that even if a backside is detectable in a theoretical model without noise, it might not be detectable in actual observations due to noise interference (manifested as either the absence of a peak corresponding to the backside or the presence of multiple fluctuations/peaks, making it difficult to discern the backside peak).

To incorporate detectability into our parameter study, we analysed the \(U_{\phi}\) observational files of IM LUP, MY Lup, and PDS453 (all of which show backsides) to calculate the standard deviation of the noise, measured in Janksy/arcsec$^2$ and averaged these standard deviations. We then converted this to Janksy/pixel to make it comparable with our integrated flux, using

\begin{equation}
\sigma_{\text{total}} = \sigma \sqrt{N}
\end{equation}

Next, we plotted the integrated flux values in a line plot, as shown in Figure \ref{fig:12}. We added noise levels at 1-sigma level ($= \sigma_{\text{total}}$), represented by the solid grey line, and the 3-sigma level ($= 3 \times \sigma_{\text{total}}$), represented by the dashed grey line, to these plots. This effectively illustrates that when the integrated flux values fall below the 3-sigma threshold, detection becomes significantly more challenging.

\subsection{Backside detection at lower inclinations}
\label{sec:5.2}
As mentioned earlier, most observed disks with a discernible backside have high inclinations, aligning with our parameter studies of tapered disks. However, there are still some disks with detectable backsides at lower inclinations. PDS 111 with an inclination of $58^\circ$ (\citealp{2024A&A...688A.149D}), IM Lup at $48^\circ$ (\citealp{2016ApJ...832..110C}), GM Aur at $55^\circ$ (\citealp{2011ApJ...732...42A}), and DG Tau at $31^\circ$ (\citealp{2010ApJ...714.1746I}) are examples of such disks.

\begin{figure}
    \centering
    \includegraphics[width=1\linewidth]{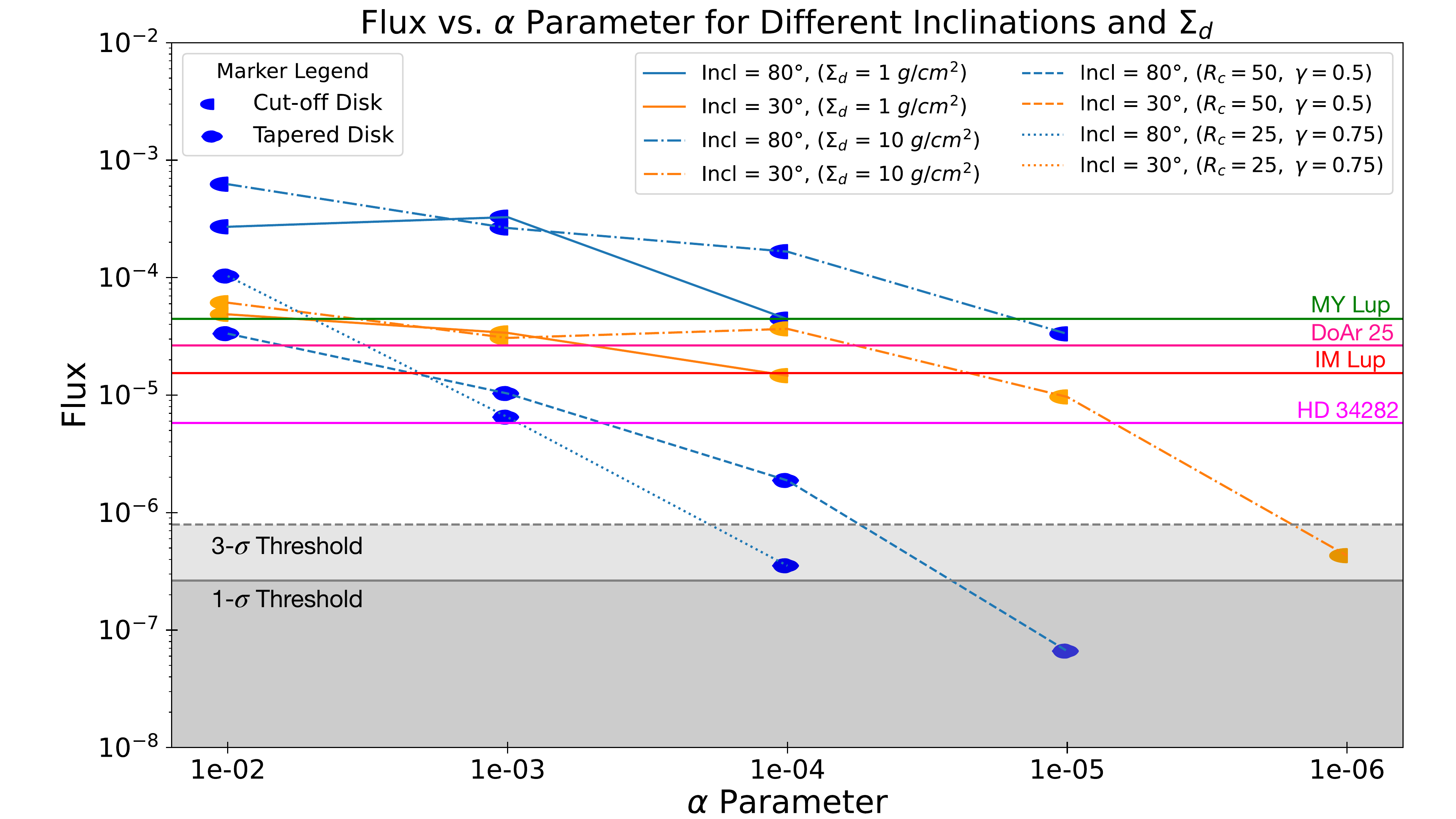}
    \caption{Line plot of integrated flux vs $\alpha$ parameter for cut-off and tapered disks at inclinations of $80^\circ$ and $30^\circ$, with flux values in jansky. The flux values are presented for cut-off disks with $\Sigma_\mathrm{d} = 1~\mathrm{g/cm^2}$ (solid lines) and $\Sigma_\mathrm{d} = 10~\mathrm{g/cm^2}$ (dash-dot lines), as well as tapered disks with $R_\mathrm{c} = 50$, $\gamma = 0.5$ (dashed lines) and $R_\mathrm{c} = 25$, $\gamma = 0.75$ (dotted lines). The blue lines represent $80^\circ$, and the orange lines represent $30^\circ$. Horizontal grey lines mark the 1-sigma (solid) and 3-sigma (dashed) noise thresholds, indicating observation sensitivity. Reference flux levels from various sources are labelled above their respective horizontal lines for clarity.}
    \label{fig:12}
\end{figure}

Examining the plots of the integrated flux of the backsides of tapered disks in Figure \ref{fig:12}, we observe that at lower inclinations, the integrated flux values are minimal or absent, as indicated by the absence of orange lines (30$^\circ$). The annotated heat map in Figure \ref{fig:4.8} shows that the integrated flux for inclinations 45$^\circ$ and 60$^\circ$ are close to or below 3-sigma detection.

First, as the inclination decreases and the disk becomes more face-on, the backside appears thinner (see the parameter study images of cut-off disks in Figure \ref{fig:PS_cut-off}). Second, the reduction in flux can be due to the presence of sufficient dust in the outer disk. The exponential decay of surface density (Equation \ref{eq:2.28}) ensures there is enough dust to attenuate the scattered light from the backside but not enough to scatter it towards the observer. The combination of these geometric effects and outer disk opacity makes detecting backsides in tapered disks more challenging.

To understand why disks with detectable backsides are still observed at lower inclinations, we examined the parameter studies of the less physical cut-off disks. The \(\alpha\) parameter, representing the Shakura-Sunyaev turbulence parameter, was set to 0.001 for tapered disk studies. However, in this comparison with cut-off disks, \(\alpha\) spans higher to lower values, similar to the approach for cut-off disks. The results show that the integrated backside flux of cut-off disks is 1–2 orders of magnitude higher than that of tapered disks. Moreover, backsides remain clearly detectable in cut-off disks even at a lower inclination of 30$^\circ$ (indicated by the orange lines), provided the disk has sufficient dust mass or turbulence.

This suggests that disks observed with a visible backside at lower inclinations might actually be truncated disks. Such truncation can result from various interactions, including close encounters with passing stars and other dynamical and tidal interactions (\citealp{2016A&A...594A..53B}, \citealp{Cuello_2018}, \citealp{2020MNRAS.491..504C}), as well as from external photoevaporation (\citealp{2018MNRAS.478.2700W}, \citealp{2022MNRAS.514.2315C} \citealp{2024A&A...681A..84G}). Flybys and other dynamical interactions can truncate the disk due to gravitational effects, particularly in regions with stellar densities exceeding $10^4$ stars per cubic parsec. Conversely, external photoevaporation, driven by high UV flux from massive stars, can strip material from the outer disk regions. Although both mechanisms typically occur in regions with higher stellar densities, their effects on the disk are distinct. Flybys can induce asymmetries and perturbations, while photoevaporation leads to a more uniform stripping of material. Nevertheless, both processes, whether combined or individual, can result in a sharp outer edge, thereby enhancing the detectability of backside flux.

\subsection{Implications for observed disks: IM Lup and PDS 111}

IM Lup and PDS 111 both exhibit well-detected backsides despite their relatively low inclinations. This observation suggests that their disks may be truncated. A sharp outer edge enhances the visibility of the backside by reducing attenuation from the extended outer regions, as discussed in Section~\ref{sec:5.2}. Various mechanisms, including external photoevaporation and dynamical interactions, can lead to such truncation, but the dominant cause likely differs between these two systems.

\citet{2017MNRAS.468L.108H} demonstrate that external photoevaporation plays a key role in shaping the outer disk of IM Lup, showing that this mechanism operates in low-mass star-forming regions and is not exclusive to more massive star-forming environments. In IM Lup, CO emission extends well beyond the millimetre continuum edge, exhibiting asymmetries that suggest active mass loss. This indicates that environmental radiation is stripping material from the outer disk, steepening the surface density profile and producing a sharply truncated outer edge. While the gaseous disk extends to nearly 1200 AU, the dust disk remains much more compact, highlighting the importance of external influences in shaping the disk. The observed truncation of the outer disk and the associated asymmetries are consistent with UV-driven mass loss, which not only limits the disk’s ability to retain gas but may also influence the detectability of the disc's backside.

PDS 111 is an intriguing case of a long-lived PPD around a solar-type star, with an estimated age of $\sim16$ Myr or potentially older. Given the age of the system, standard disk evolution models predict significant dispersal, yet PDS 111 still retains a gas-rich disk with a young appearance. \citet{2024A&A...688A.149D} suggest that external photoevaporation is unlikely to be a dominant factor in this system, as it would have led to the rapid dissipation of the disk. The relative isolation of the system in the foreground of the Orion cloud further supports this, as it lacks the close stellar neighbours that are typically required for strong external photoevaporation.

Interestingly, scattered light and ALMA observations reveal a warp or asymmetry in the outer disk, along with possible misalignment in the inner regions traced by variable H$\alpha$ emission. A possible explanation is an external perturber or past close encounter, which may have also truncated the disk, affecting its overall structure and making the backside of the disk visible in scattered light. If a perturber were responsible for both the truncation and the observed asymmetry, it could provide a self-consistent scenario for PDS 111’s extended disk lifetime while explaining its structural features and backside visibility.

\subsection{Optical depth}

\begin{figure}
    \centering
    \includegraphics[width=1\linewidth]{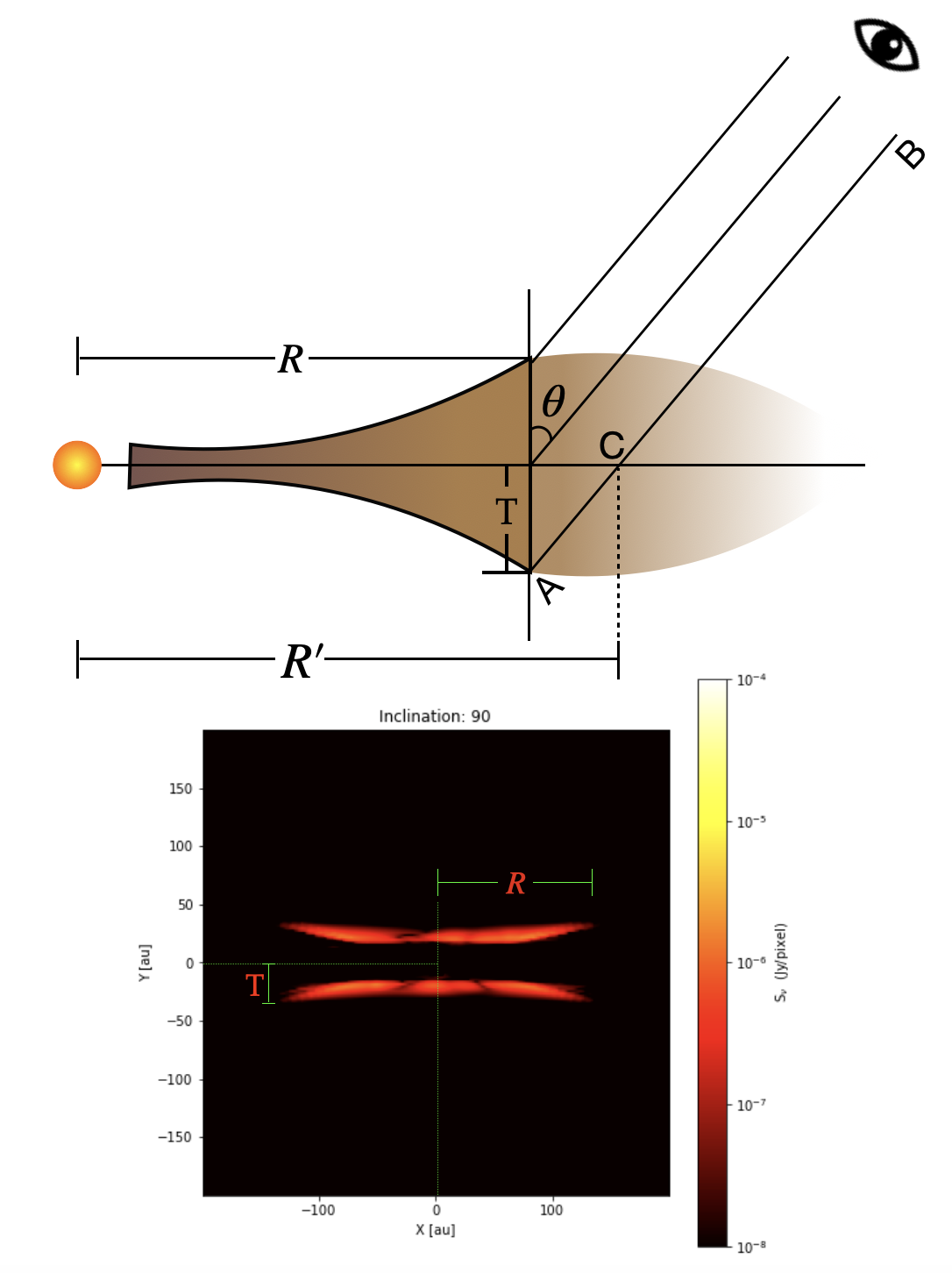}
    \caption{Top: Geometry of the disk used for optical depth calculations. Bottom: Representation of how $R$ and $T$ are inferred from the simulation of an edge-on disk.}
    \label{fig:geometry-sketch}
\end{figure}

\begin{figure}
    \centering
    \includegraphics[width=1\linewidth]{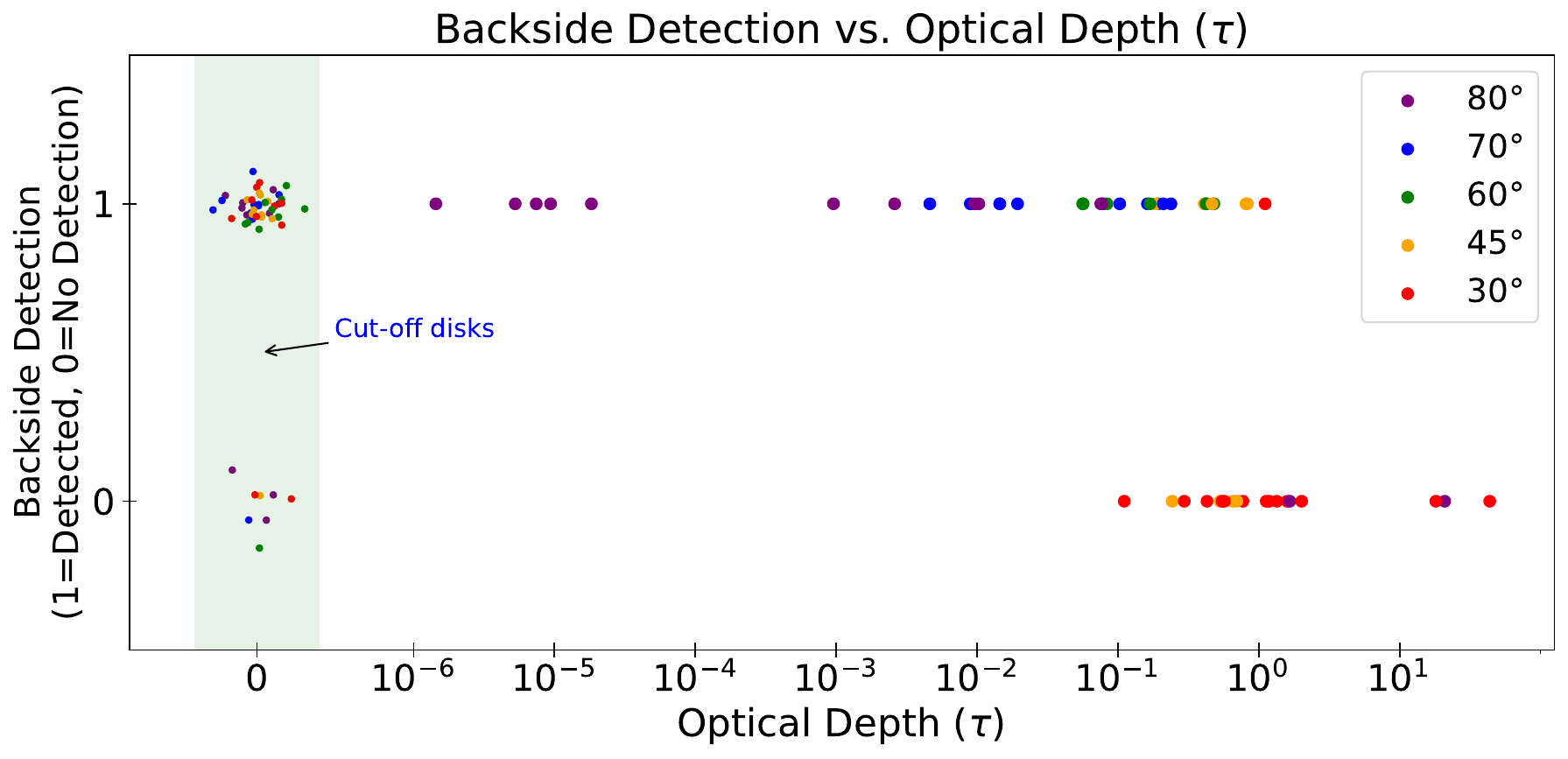}
    \caption{Plot of optical depth versus backside detectability for cut-off and tapered disks at various inclinations. A slight jitter is applied to the cut-off disk data points in both the x and y axes to improve visibility, as the optical depth of all points corresponding to cut-off disks is zero, causing them to overlap if unadjusted. The entire green bar represents an effective optical depth of 0, corresponding to cut-off disks.}
    \label{fig:detection-tau}
\end{figure}

The optical depth, \( \tau \), of a medium quantifies the attenuation of radiation as it traverses through the material. At each location r in the disk, the vertical optical depth can be expressed as
\begin{equation}
\tau(r) = \rho_\mathrm{d}(r, z) \cdot \kappa \cdot \int_{-\infty}^{\infty} dz,
\end{equation}
where \( \rho_\mathrm{d}(r, z) \) is the dust volume density (mass per unit volume), \( \kappa \) is the opacity (cross-section per unit mass) at the relevant wavelength, and \( \int_{-\infty}^{\infty} dz \) represents the path length along the vertical direction through the medium. The surface density of dust, \( \Sigma_\mathrm{d}(r) \), is defined as
\begin{equation}
\Sigma_\mathrm{d}(r) = \int_{-\infty}^{\infty} \rho_\mathrm{d}(r, z) \, dz,
\end{equation}
allowing the optical depth to be rewritten as
\begin{equation}
\tau(r) = \kappa_{\text{total}} \cdot \Sigma_\mathrm{d}(r).
\end{equation}
Here, \( \kappa_{\text{total}} \) is the total opacity, which incorporates contributions from multiple dust species. It is computed as
\begin{equation}
\kappa_{\text{total}} = \sum_i f_i \cdot \kappa_i,
\end{equation}
where \( f_i \) represents the mass fraction of each dust species, and \( \kappa_i \) denotes the corresponding opacity.

For an inclined disk, the path length through the medium increases due to the geometry of the line of sight. Accounting for the inclination angle \( \theta \), the effective optical depth is therefore modified by a factor of \( 1/\cos \theta \). The inclination-corrected optical depth is given by
\begin{equation}
\tau_{\text{incl}} = \kappa_{\text{total}} \cdot \Sigma_\mathrm{d}(r) \cdot \frac{1}{\cos \theta}.
\end{equation}

To analyse the detection of the disc's backside, consider the line of sight passing through the disk along the path \( \text{AB} \), which intersects the midplane at point \( C \). The distance \( R' \), from the star to point \( C \), needs to be determined analytically. This requires estimates of the disk radius, \( R \), and thickness, \( T \). These quantities can be constrained using edge-on simulations of the disk, assuming similar physical parameters to those in the observed case. The geometry is provided in figure \ref{fig:geometry-sketch}. Using trigonometry, \( R' \) can be expressed as
\begin{equation}
R' = R + T \cdot \tan(\theta).
\end{equation}

It is important to note that \( R \) is not the characteristic radius (\( R_\mathrm{c} \)) from the truncated disk equation, but rather the distance up to which the disk is observed, or the point at which we assume the backside of the disk becomes visible. This assumption is based on the visibility of the front side up to \( R \) within the selected intensity range. Furthermore, it is not necessary for the exponential tapering to commence precisely at \( R \). 

Given the surface density of the disk at point \( C \), \( \Sigma_\mathrm{d}(R') \), the optical depth along the inclined line of sight can be approximated by
\begin{equation}
\tau_{\text{AB}} = \kappa_{\text{total}} \cdot \Sigma_\mathrm{d}(R') \cdot \frac{1}{\cos \theta}.
\end{equation}

In the case of cut-off disks, the disk material is abruptly truncated at the radius \( R \), meaning there is no mass distribution beyond this boundary. Consequently, the surface density at \( R' \) becomes \( \Sigma_\mathrm{d}(R') = 0 \). As a result, the optical depth along the line of sight, \( \tau_{\text{AB}} \), is effectively zero (\( \tau_{\text{AB}} = 0 \)). This implies that there is no intervening material in the outer disk to obstruct the view of the disk’s backside.

We explore the relationship between optical depth (\( \tau \)) and the detectability of the disc's backside in figure \ref{fig:detection-tau}. Optical depth is plotted on the x-axis, while the y-axis represents the backside detectability, where \( y = 1 \) indicates detectable and \( y = 0 \) indicates non-detectable.

For cut-off disks, the optical depth along the line of sight is zero, as no material exists beyond the truncation radius \( R \). As a result, all data points for cut-off disks are concentrated in the vertical green region at \( \tau = 0 \). To enhance visualisation, a slight jitter is applied to these points along both the \( x \)- and \( y \)-axes. This jitter allows for a clearer differentiation between the points corresponding to different inclinations. Despite the optical depth being effectively zero for cut-off disks, the jitter ensures that the data points are distinguishable. Points near \( y = 1 \) correspond to detectable backside, while those near \( y = 0 \) represent non-detectable backside.

For tapered disks, the visualisation is more straightforward between optical depth and backside detectability. When \( \tau \) is less than \( 0.1 \), the backside is detectable, indicated by points along \( y = 1 \). As \( \tau \) increases beyond \( 1 \), the backside becomes obscured, with points along \( y = 0 \). 

In the intermediate range, where \( 0.1 < \tau < 1 \), the detectability is mixed. Higher inclination angles, such as \( 80^\circ \), allow for backside detection even at higher optical depths, while lower inclinations, such as \( 30^\circ \), result in near-zero detectability. This threshold highlights the combined influence of inclination angle and optical depth on the visibility of the disc's backside. Higher inclinations allow for greater backside detection across a wider range of optical depths, whereas lower inclinations result in diminished visibility, particularly as the optical depth increases.

\subsection{Future work}
\label{sec:5.3}
The next phase of our study involves validating our findings through comparison with observational data to determine if the integrated flux values we derived correspond to observed backside features. These flux values can serve as reference points for future observations, helping us assess whether a backside may be observable and if the detection is significant enough to be considered robust.

Our study focused exclusively on axisymmetric disks without additional structures, whereas real disks exhibit a variety of features such as rings, spirals, gaps, and warps (\citealp{2018A&A...620A..94G}). Additionally, real disks often contain ambient material and planetary bodies, which can introduce complexities not accounted for in our models. Flybys and other dynamic interactions can lead to misalignments, creating broad or narrow shadows that influence backside visibility (\citealp{2020A&A...635A.121M}). Investigating how these structural elements affect the detectability of backside features in scattered light imaging is essential for refining our models and improving our understanding of disk properties.

While our study utilised standard DIANA model dust opacities with amorphous pyroxene (70\% Mg), actual PPDs may contain grains with varying chemical compositions in their outer regions. Investigating different material compositions, such as icy or more porous dust, including water ice, is interesting for assessing their impact on backside visibility. Additionally, it would be beneficial to examine whether the observed disks with lower inclinations, where backside features are detected, are indeed truncated by the mechanisms described earlier.

\section{Conclusion}
\label{section_6}

Scattered light imaging, particularly with PDI, has become essential in exploring protoplanetary disks. These disks comprise two distinct scattering surfaces: the front side, oriented towards the observer, and the backside, oriented away from the observer. The prominence of the backside is apparent when the observer's line of sight intersects with the forward-scattered light from the backside of the disk. Studying these backsides provides a 3D perspective on the disk structure and dust properties. Conversely, the absence of discernible backside features may be due to optically thick regions, unfavourable inclination angles, material deficits, or observational noise. Understanding these conditions refines our knowledge of disk dynamics and impacts data interpretation, aiding in model constraints and hypotheses related to disk evolution.

To further understand backside detectability, we used the radiative transfer code \texttt{RADMC-3D} to simulate scattered light images of disks and conduct parameter studies. Our goal was to determine the factors influencing backside visibility and why some disks show a visible backside while others do not.

The main conclusions are:

\begin{itemize}
    \item The tapered disk model with an exponential taper, accounting for viscous spreading, can effectively hide the backside, aligning with observations where only a few disks show discernible backsides.
    \item Disks with visible backsides at low inclinations could indeed be cut-off (truncated) disks, as detecting backsides in tapered disks at low inclinations is harder and rarer.
    \item Factors such as total matter in the outer disk, mass distribution among dust grains, turbulence, and vertical stratification (settling) in the outer disk also affect backside visibility. 
\end{itemize}

\begin{appendix}
\section{Model images from parameter studies}
This appendix contains the images of additional parameter studies
executed for the paper. For a discussion of the figures, see the main
text of this paper.
    
\begin{figure*}[h!]
    \centering
    \includegraphics[width=\textwidth]{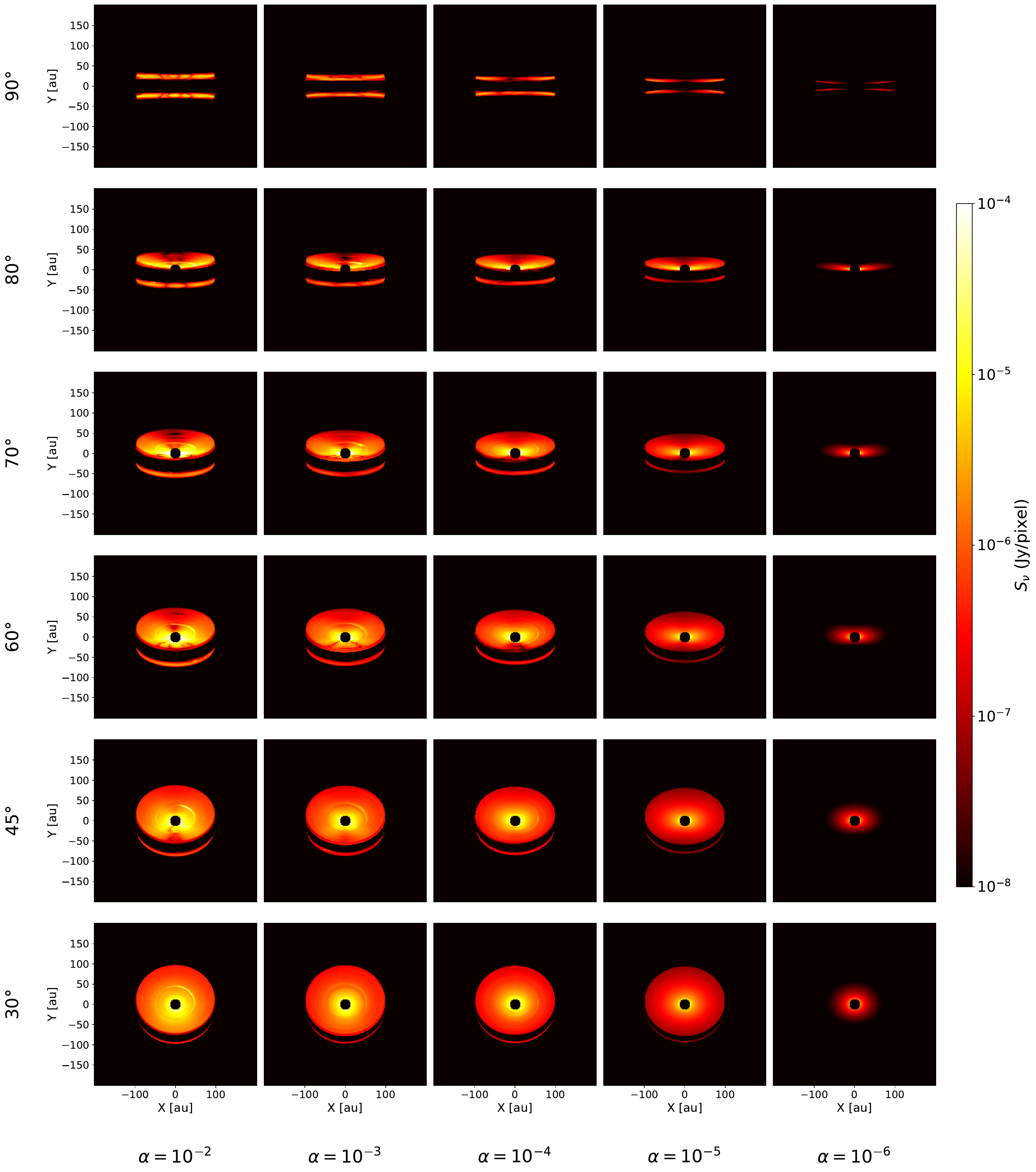}
    \caption{Parameter study of cut-off disks with $\Sigma_\mathrm{d}$ = 10 g/cm$^2$ at $R_\mathrm{c} = 1$ AU for varying the Shakura-Sunyaev turbulence parameter ($\alpha$). Each column represents observations of the disk at different inclinations for a specific $\alpha$ value, with corresponding $\alpha$ values listed on the x-axis. Note that the turbulence parameter $\alpha$ decreases from left to right, indicating reduced turbulence in the images towards the right.}
    \label{fig:PS_cutoff_1000}
\end{figure*}

\begin{figure*}
    \centering
    \includegraphics[width=\textwidth]{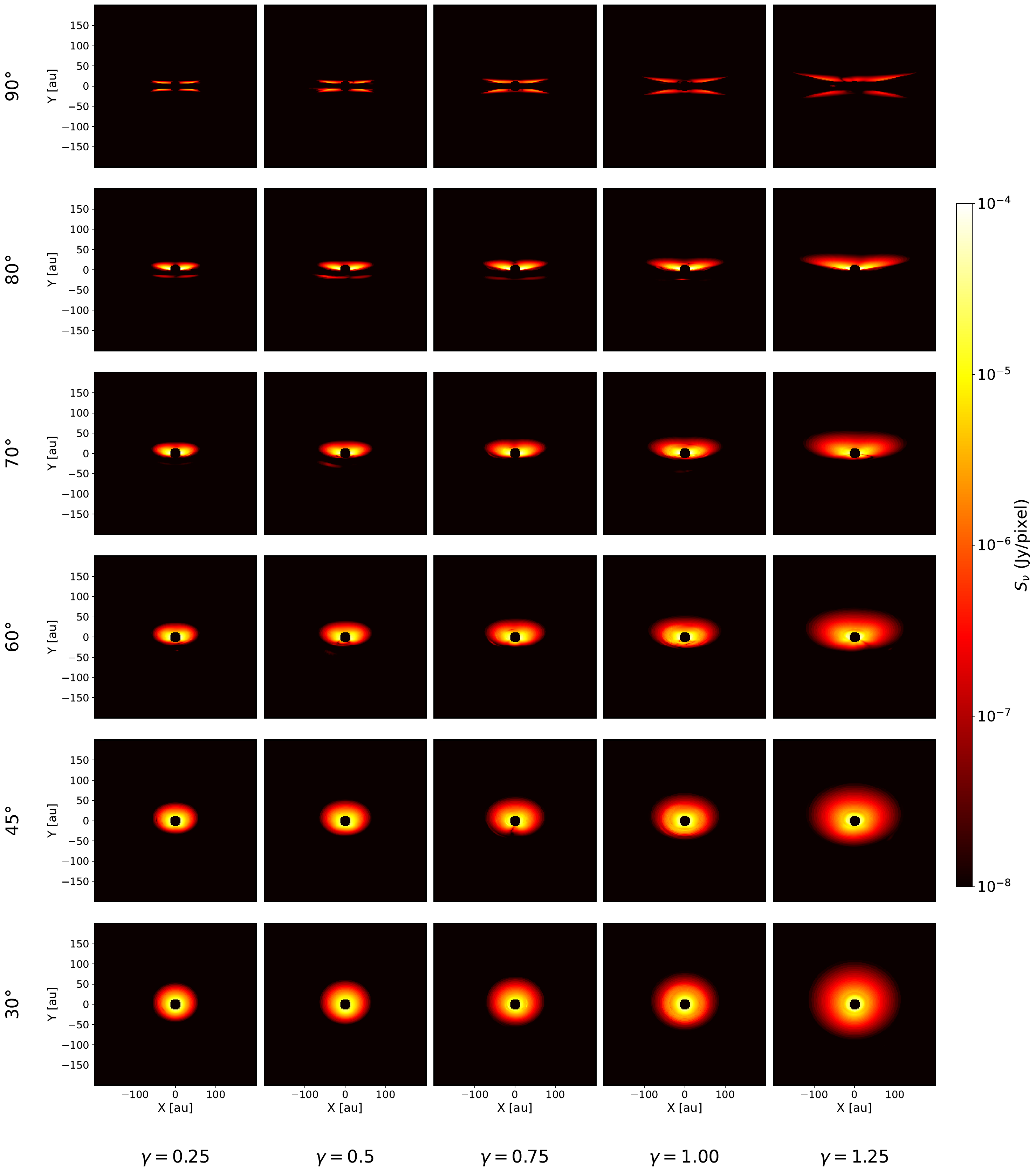}
    \caption{Parameter study of tapered disks with a varying negative power-law exponent ($\gamma$), with a fixed characteristic radius, $R_\mathrm{c}$ = 25 AU, and a Shakura-Sunyaev turbulence parameter, (\(\alpha\)) = 0.001. Each column represents observations of the disk at different inclinations for a specific $\gamma$ value, with corresponding $\gamma$ values listed on the x-axis.}
\end{figure*}

\begin{figure*}
    \centering
    \includegraphics[width=\textwidth]{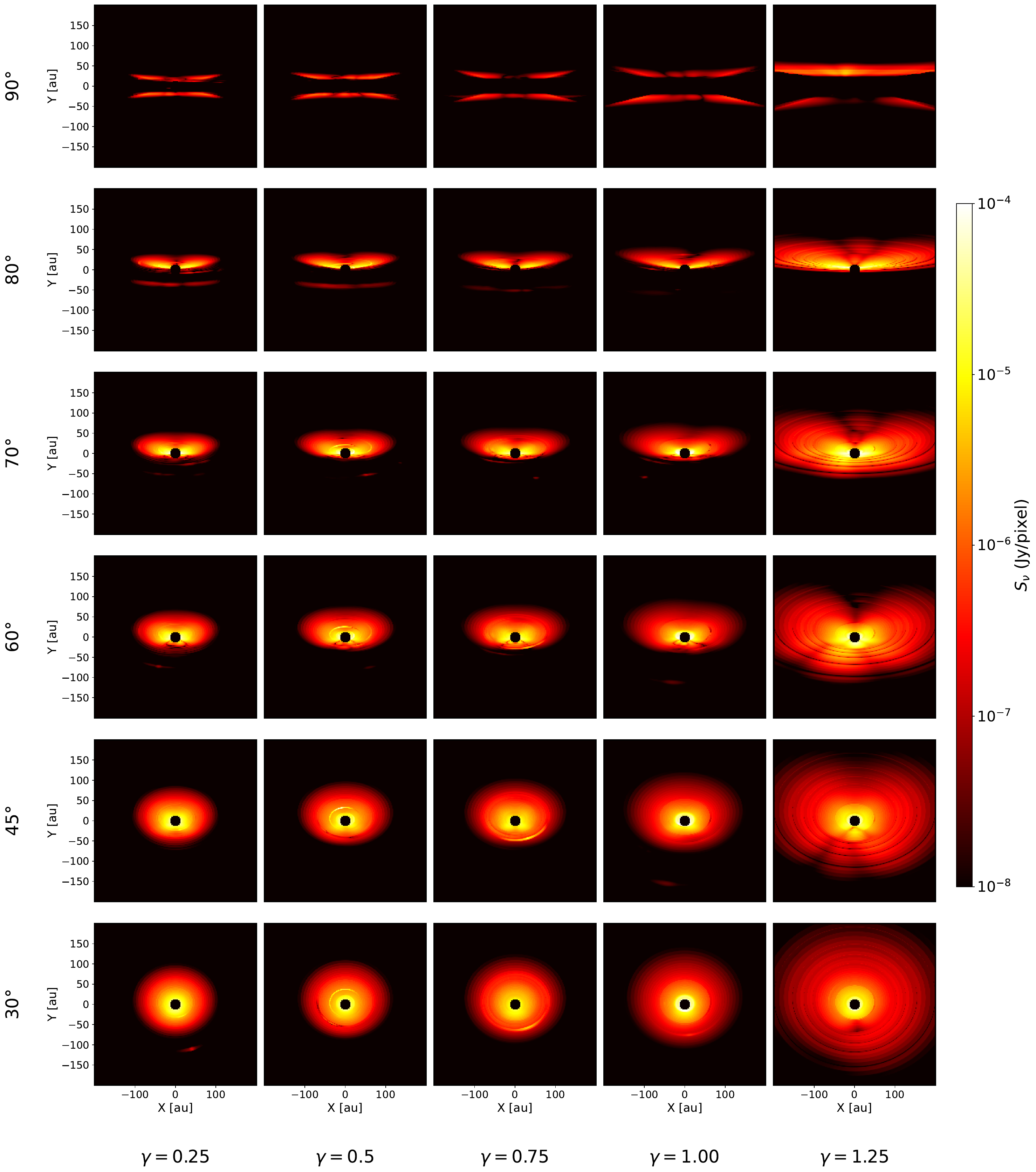}
    \caption{Parameter study of tapered disks with varying negative power-law exponent ($\gamma$), with a fixed characteristic radius, $R_\mathrm{c}$ = 50 AU, and a Shakura-Sunyaev turbulence parameter, (\(\alpha\)) = 0.001. Each column represents observations of the disk at different inclinations for a specific $\gamma$ value, with corresponding $\gamma$ values listed on the x-axis.}
    \label{fig:combined_tapered}
\end{figure*}

\begin{figure*}
    \centering
    \includegraphics[width=1\linewidth]{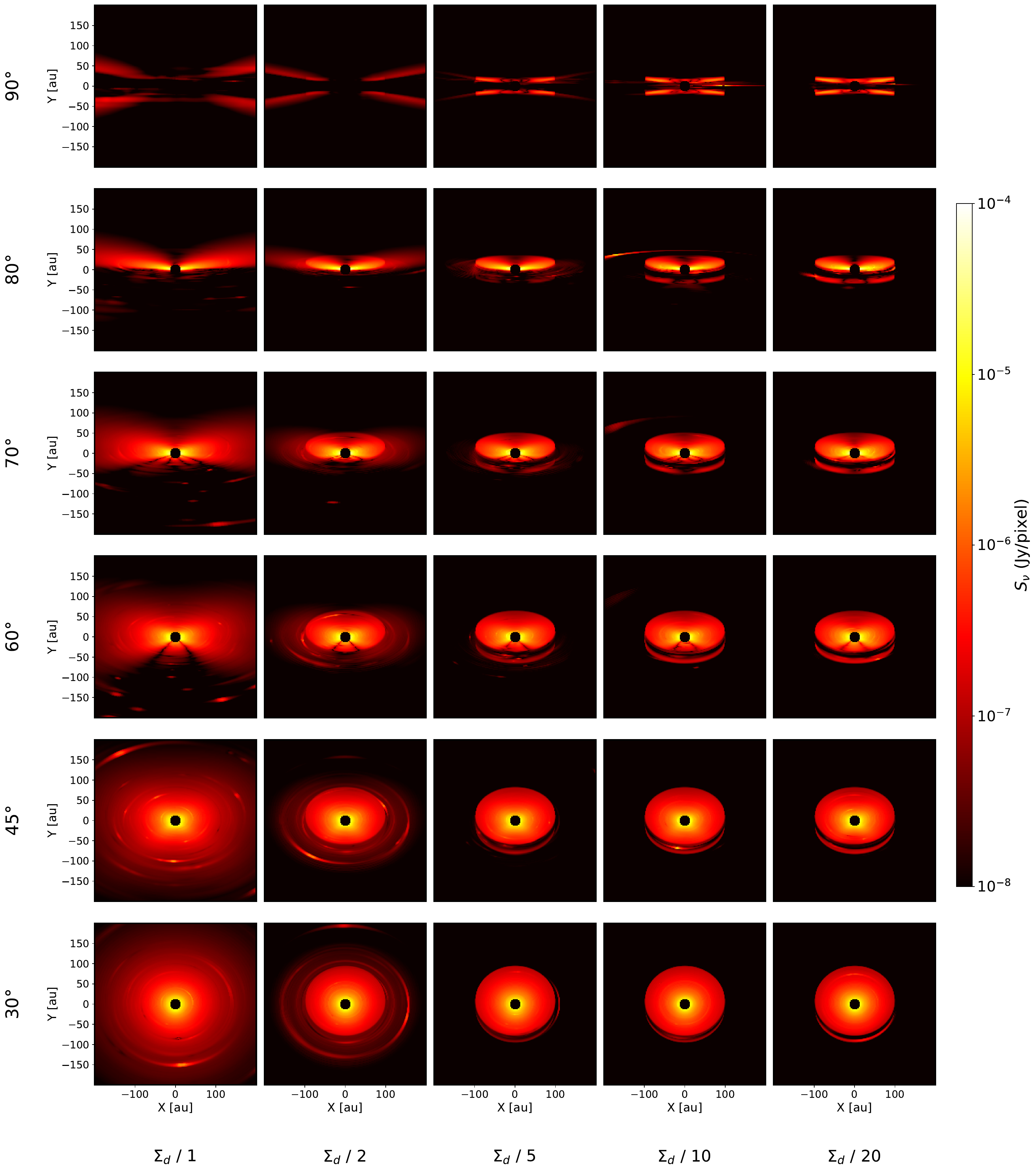}
    \caption[Parameter study of extended disk images with $\Sigma_\mathrm{d} = 1$ g/cm$^2$ at $R_\mathrm{c} = 1$ AU and $\alpha = 0.001$ for varying the reduction factors of surface density ($\Sigma_\mathrm{d}$) in the outer disk.]{Parameter study of extended disk images with $\Sigma_\mathrm{d} = 1$ g/cm$^2$ at $R_\mathrm{c} = 1$ AU and Shakura-Sunyaev turbulence parameter $\alpha = 0.001$. The disk extends from 100 AU to 500 AU. Each column represents observations of the disk at different inclinations, with corresponding reduction factors of surface density ($\Sigma_\mathrm{d}$) in the outer disk shown in the last row.}
    \label{fig:4.9}
\end{figure*}

\end{appendix}

\end{document}